\documentclass[prd,twocolumn,showpacs,floatfix,amsmath,nofootinbib,amssymb,floatfix]{revtex4}
\usepackage{graphicx,color,dcolumn,booktabs,bm}
\usepackage{longtable,lscape}
\usepackage{txfonts}
\usepackage{overpic}
\usepackage{amssymb}
\usepackage{indentfirst}
\usepackage{feynmf}   %{feynmp}
\usepackage{slashed}  %for Feynman symbols
\usepackage{cases}
\usepackage{color}
\usepackage{multirow}
\usepackage{epstopdf}
\usepackage{longtable}
\usepackage{graphicx,color,dcolumn,booktabs,bm}
\usepackage[colorlinks,
            citecolor=blue,
            anchorcolor=red,
            menucolor=red,
            linkcolor=red,
            filecolor=red,
            runcolor=red,
            urlcolor=blue,
            frenchlinks=red]{hyperref}

\graphicspath{{Figures/}} %

\allowdisplaybreaks

\begin{document}
%\begin{CJK}{GBK}{}

\title{Universality and Regge-like spectroscopy for orbitally-excited light mesons}

\author{Duojie JIA}
\email{jiadj@nwnu.edu.cn}
\author{Wen-Chao Dong}

\affiliation{Institute of Theoretical Physics, College of Physics and
Electronic Engineering, Northwest Normal University, Lanzhou 730070, China}

\begin{abstract}
A new Regge-like mass relation for excited light mesons is presented in
relativized quark model which supports an universality that the quark mass
dependence of the light meson spectroscopy is suppressed significantly and the
confining parameter is nearly family independent. It is obtained by using
auxiliary field method and quasi-linearizing the solution to the mass relation
solved from the model. The resulted mass predictions are in good agreement
with the observed masses for the orbitally-excited trajectory family of
$\pi/b$, $\rho/a$, $\eta/h$, $\omega/f$, $K^{\ast}$ and $\phi/f^{\prime}$. A
semiclassical argument is given that the inverse slopes on the radial and
angular-momentum Regge trajectories are equal in the massless limit of quarks.
\end{abstract}


\maketitle

\section{Introduction}\label{sec1}

{{The dynamics behind the formation of light hadrons still remains to be unclear
four decades after the discovery of the theory of strong interaction, the
quantum chromodynamics (QCD). In the case of the light mesons (which we shall
discuss in this work) that composed of light quarks ($n\equiv(u,d),s$), the
most lowest state are well established (with some exceptions, the $0^{++}$
scalar mesons etc.), while the excited states in the range $M>2GeV$ are less
understood. Despite difficulties in solving QCD exactly, it is expected that
properties and decay of light mesons will shed light on understanding of QCD
at low energy. Recently, advances in experiment \cite{Patrignani:C16}, with
light mesons generated in copious amounts, make it possible to address such
issues as whether nonconventional states (exotics) exist in the light sector
of hadrons. For instance, $a_{0}(980)$ and $f_{0}(980)$ were expected to be
exotic \cite{WeinsteinI:1982,WeinsteinI:1983}. However, it is fair to say that
we may not truly understand the spectrum of light mesons before we understand
the excitations of the lowest mass mesons. Further, the light-meson spectrum
has become important not merely for the intrinsic understanding of these
states, but also as a prerequisite for exploring exotic states (see
\cite{GodfreyN:P1999,AmslerT:RP04} for a review).

On the other hand, the observed spectrum of light mesons manifests themselves
in almost universal pattern: they populate approximately linear Regge
trajectories \cite{ChewF:61}, almost parallel between trajectories
\cite{AnisovichAS:PRD00,MasjuanAB:D12}. This universal pattern of the hadron
spectrum strongly hints that the formation dynamics of light mesons is more
or less universal by itself in the sense that their spectrum are almost
independent of the quark favors, as QCD is. Great success has been achieved in
building dynamical quark models to describe the whole meson spectrum in an
universal way (see,
\cite{RujulaGG:D75,KangS:D75,BasdevantB:C85,GodfreyIsgur:85,EbertFG:D09} for
instance) and argument was given \cite{Olsson:D1997} that universality can
arise from the relativistic effects and the confinement dynamics. In the high
excitation spectrum, however, this feature remains to be understood yet.

For light hadrons, a remarkable feature of Regge trajectory is that the slope
$\alpha^{\prime}$ in the Chew-Frautschi relation, $L=\alpha^{\prime}%
M^{2}+\alpha(0)$, where $L$ is the angular momentum of the hadron state and
$M$ its mass, depends weakly on the flavor content of the states lying on the
corresponding trajectory. The Regge slope $\alpha^{\prime}$ varies slightly
from trajectory to trajectory, by less than $10\%$ for nonstrange mesons
\cite{AnisovichAS:PRD00,MasjuanAB:D12,Sharov:13,SonnenscheinW:JHEP14}, and the
linearity of trajectories are commonly assumed. When strangeness involved, the
situation becomes slightly involved. Nonlinearity in Regge trajectories was
suggested \cite{BrisudovaBG:00} and the correction to the linear trajectory are
explored in Refs. \cite{JohnsonN:PRD79,Afonin:B03} and Ref.
\cite{SonnenscheinW:JHEP14}, for instance. It is then of important to examine
carefully the properties of Regge trajectory with enriched experimental data
of the light mesons. The knowledge of Regge trajectories is also valuable in
the recombination and fragmentation models for hadrons transition in the
scattering region ($t<0$) \cite{Collons:77}.

Purpose of this work is to explore the universality high in the excited
spectrum of the light mesons using relativistic quark model combined with
auxiliary field method. We propose a new Regge-like mass relation for the
orbitally-excited light mesons which supports a universality that the quark
mass dependence of the light meson spectroscopy is suppressed. The obtained
mass relation is tested against the observed masses of mesons considered. It
is found that the parameters in the relation are roughly universal except the
vacuum constant. A explicit expression for the Regge slope and intercept was
obtained and compared to the results extracted from the other analysis of the
meson family of $\pi/b$, $\rho/a$, $\eta/h$ and $\omega/f$ in the ($L$,$M^{2}$)
planes or predictions in the literatures. Suggestion is made that the members
of the $\eta/h$ trajectory may contain components of exotics.

We also discuss the implications of our results in comparison with that in
the string (flux-tube) picture of mesons \cite{Nambu:D74,Goto:71} and that in
other quark models. By the way, semiclassical argument is given that the
slopes on the radial and angular-momentum Regge trajectories are equal in the
massless limit of quarks, as suggested by Anisovich et
al. \cite{AnisovichAS:PRD00}, Afonin \cite{Afonin:07,Afonin:BA06},
Bicudo \cite{Bicudo:D07} and Forkel et al. \cite{Forkel:07a,Forkel:07b}. In the
latest case, the meson spectrum is predicted to be
\cite{Forkel:07a,Forkel:07b}
\begin{equation}
M^{2}=4\lambda^{2}(n+L+1/2), \label{MJ}%
\end{equation}
with $n$ the radial quantum number of the state. For more discussions of the
Regge-like relation, see
\cite{KangS:75,MaungKN:D93,VeseliO:B96,Afonin:BA06,Bicudo:D07,Afonin:07,SilvestreSB:JPA2008}
for instance.

\section{The light quark dynamics and auxiliary fields}\label{sec2}

We begin with the dynamics of the relativized quark model
\cite{Durand:D82,LichtenbergNP:pl82,Durand:D84,GodfreyIsgur:85,Jacobs:D86}
with the spin-dependent interactions ignored. It is given by the spinless
Salpeter Hamiltonian
\begin{equation}
H_{0}=\sum_{i=1}^{2}\sqrt{\mathbf{p}_{i}^{2}+m_{i}^{2}}+V, \label{SH}%
\end{equation}
where $\mathbf{p}_{i}$ ($\mathbf{p}_{1}=-\mathbf{p}_{2}=\mathbf{p}$) is the
particle momentum of the quark $i$, and $V$ the interquark interaction
given by the usual linear confining potential $ar$ plus the short-range
color-Coulomb potential $-k_{s}(r)/r$,%
\begin{equation}
V=ar-\frac{k_{s}(r)}{r}+V_{0}. \label{V}%
\end{equation}
\ Here $k_{s}(r)=4\alpha_{s}(r)/3$, with $\alpha_{s}(r)$ the strong coupling,
defined as the Fourier transformation of the running QCD coupling $\alpha
_{s}(Q^{2})$, which depends on the relative coordinate $r=|\mathbf{x}%
_{1}-\mathbf{x}_{2}|$ of the quark $1$ and antiquark $2$ with the bare masses
$m_{1}$ and $m_{2}$, respectively. The mass $m_{i}$ are that of bare quarks,
$m_{i=u,d}=3$MeV and $m_{i=s}=96$MeV, which differs our approach from most of
the quark models with $m_{i}$ the valence quark masses, see Section 5 for
discussions. As emphasized in Ref. \cite{LuchaRS:D92}, $V_{0}$ is a parameter
as fundamental and indispensable as the quark masses and slope of the linear
potential $a$. For the lattice evidences for the interaction (\ref{V})
in the heavy-flavor sector, see \cite{Bali01,KawanaiS:12}.

At short distance (high energy), the coupling $\alpha_{s}$ in (\ref{V})
depends on energy scale $Q$ along the renormalization group equations in a
known way \cite{Chekanov:B03,Bethke:G00}. At long distance(or in the infrared
region), the actual value of $\alpha_{s}$ at a given $Q$ relies mainly on
experiment \cite{Patrignani:C16}, and remains to be
explored \cite{ShirkovSol:97,Ganbold:D10,ALPHA coll:17}. The authors of
Ref. \cite{GodfreyIsgur:85} use a functional (sum of $e^{-Q^{2}/4s_{k}}$) to
mimic the running of $\alpha_{s}(Q^{2})$ and its possible saturation
\cite{BrodskyT:B04,AguilarMN:A04,ShirkovSol:97,Ganbold:D10} at some critical
value $\alpha_{s}^{critical}=\alpha_{s}(Q^{2}\rightarrow0)$(the infrared fixed
point) when $Q^{2}$ becomes low and the confinement emerges. Written in the
position space, this functional has the $\operatorname{erf}$ form
\begin{equation}
\alpha_{s}(r)=\sum_{k=1}^{3}\alpha_{k}\operatorname{erf}(s_{k}r),
\label{strong}%
\end{equation}
where $\operatorname{erf}(x)$ is the error function, $\alpha_{k}%
=\{0.25,0.15,0.2\}$, $s_{k}=\{1/2,\sqrt{10}/2,\sqrt{1000}/2\}$, and
$\alpha_{s}^{critical}=\sum_{k=1}^{3}\alpha_{k}=0.6$. A nontrivial IR-fixed
point around $\alpha_{s}(\infty)=0.7$ is suggested recently with respect to
the confinement scale $\Lambda=345$MeV \cite{Ganbold:D10}.

Lacking adequate knowledge of the strong coupling, we approximate, for
simplicity, the color-Coulomb interaction in (\ref{V}) by
\begin{equation}
\frac{4}{3}\frac{\alpha_{s}(r)}{r}\simeq\frac{k_{\infty}}{r+\lambda/\Lambda},
\label{ascut}%
\end{equation}
which well fits the color-Coulomb interaction in the long-distance region, as
shown in FIG. 1. Here, $k_{\infty}=4\alpha_{s}(\infty)/3=0.8$ corresponds to
$\alpha_{s}(\infty)=0.6$ used in Ref. \cite{GodfreyIsgur:85}. The deviation
produced by the approximation (\ref{ascut}), $Err=\left\langle k_{s}%
(r)/r-k_{\infty}/(r+\lambda/\Lambda)\right\rangle _{\Phi}$, was listed
explicitly in Table I(a) for the weighted function of the harmonic oscillator
$\Phi=R_{nl}(r)$ and $\lambda=0.1$.
%TCIMACRO{\FRAME{ftbpFU}{3.6227in}{2.3877in}{0pt}{\Qcb{The comparison of the
%Coulomb interquark-interaction with the regularized Coulomb potential}}%
%{}{fitcolor-colombcut1-301.eps}{\special{ language "Scientific Word";
%type "GRAPHIC";  maintain-aspect-ratio TRUE;  display "USEDEF";
%valid_file "F";  width 3.6227in;  height 2.3877in;  depth 0pt;
%original-width 3.8099in;  original-height 2.5007in;  cropleft "0";
%filename 'FitColor-Colombcut1-301.eps';file-properties "XNPEU";}} }%
%BeginExpansion
\begin{figure}
[ptb]
\begin{center}
\includegraphics[
width=3.4in
]%
{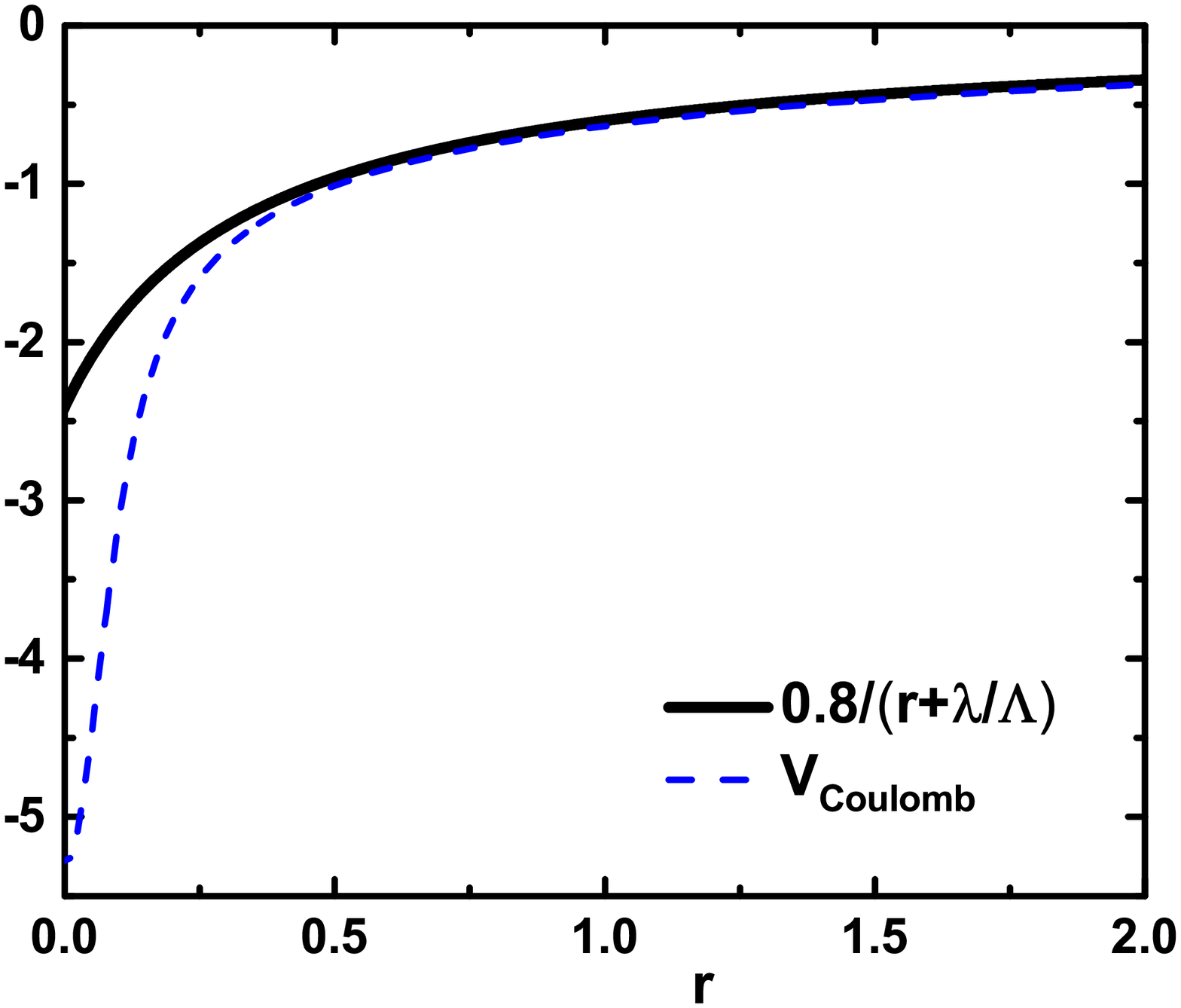}%
\caption{Comparison of the color-Coulomb interaction with the
regularized Coulomb potential.}%
\end{center}
\end{figure}
%EndExpansion
\medskip

\renewcommand\tabcolsep{0.13cm}
\renewcommand{\arraystretch}{1.8}
\renewcommand{\raggedright}{\leftskip=0pt \rightskip=0pt plus 0cm}
\begin{table}[!htbp]
\flushleft
{{\small TABLE I(a): The averaged deviation $Err=\left\langle k_{s}(r)/r-k_{\infty
}/(r+\lambda/\Lambda)\right\rangle $ caused by the approximation (\ref{ascut})
of the color-Coulomb interaction for $k_{\infty}=0.8,\lambda=0.1$ and
$\Lambda=0.301$GeV. The weighted function is that of the harmonic oscillator
$\Phi=R_{nl}(r)$ with the harmonic oscillator length $a_{H}=0.18$.}}%

\begin{tabular}
[c]{cccccccc}\hline\hline
$l$ & $0$ & $1$ & $2$ & $3$ & $4$ & $5$ & $6$\\\hline
$Err(10^{-2})$ & $-2.59$ & $-1.93$ & $-1.49$ & $-1.18$ & $-0.96$ & $-0.81$ &
$-0.69$\\\hline\hline
\end{tabular}
\medskip
\end{table}

To explore the orbitally-excited spectrum of the light mesons, we extend the
analysis in Ref. \cite{JIAPH:ijmpa17} to the case of massive strange quark:
$m_{s}=96$MeV. Following \cite{SemaySN:2004,JIAPH:ijmpa17}, we employ the
auxiliary field (AF)
method \cite{AMPolyakov:87,GubankovaYD:PLB94,MorgunovNS:PLB99} to formally
enlarge problem of the Hamiltonian (\ref{SH}) to a family of Hamiltonians
parameterized by three auxiliary fields \{$\mu_{1},\mu_{2},\nu$\} and solve
them in the enlarged Hilbert space. The eigenvalues of the Hamiltonian
(\ref{SH}) follows from the parameterized energy levels by shrinking the
parameterized space back to the original space. The point is to employ the
relation $\sqrt{B}=\min_{\lambda}\{\frac{B}{2\lambda}+\frac{\lambda}{2}\}$
(the minimization achieved when $\lambda=\sqrt{B}>0$) to reformulate the
Hamiltonian (\ref{SH}) as $H=\min_{\mu_{1,2},\nu}\left\{  H(\mu_{1,2}%
,\nu)\right\}  $, where
\begin{align}
H(\mu_{1,2},\nu)  &  =\sum_{j=1}^{2}\left[  \frac{\mathbf{p}^{2}+m_{j}^{2}%
}{2\mu_{j}}+\frac{\mu_{j}}{2}\right]  +\frac{a^{2}r^{2}}{2\nu}+\frac{\nu}%
{2}\nonumber\\
&  -\frac{k_{s}(r)}{r}+V_{0}, \label{Hmn}%
\end{align}
with the auxiliary fields $\{\mu_{1},\mu_{2},\nu\}$ being operators
quantum-mechanically. These fields has to be eliminated as the Lagrange
multipliers eventually. One can show that $H(\mu_{1,2},\nu)$ is equivalent to
(\ref{SH}) up to the elimination of $(\mu_{1},\mu_{2},\nu)$ through the
constraints%
\begin{align}
\delta_{\mu_{j}}H(\mu_{1,2},\nu)  &  =0\Longrightarrow\mu_{j}\rightarrow
\mu_{j,0}=\sqrt{\mathbf{p}_{j}^{2}+m_{j}^{2}},\nonumber\\
\delta_{\nu}H(\mu_{1,2},\nu)  &  =0\Longrightarrow\nu\rightarrow\nu
_{0}=a|\mathbf{x}_{1}-\mathbf{x}_{2}|=ar. \label{min}%
\end{align}

Assuming that the quantum average of the AFs $\langle\mu_{i,0}\rangle
\mathbf{\gg}m_{i}$ ($i=1,2$), which is the case for the light quarks in the
excited mesons, for which the averaged momentum $\langle\mathbf{p}_{j}%
^{2}\rangle\mathbf{\ }$is large enough compared to the bare masses $m_{i}$,
one can view, using the Born-Oppenheimer approximation, the average
$\langle\mu_{i,0}\rangle$ $=\langle\sqrt{\mathbf{p}_{i}^{2}+m_{i}^{2}}\rangle$
as slow variables, being the effective dynamical mass of the quark $i$, and
thereby treat them as real $c$-numbers \cite{SilvestreSB:JPA2008}. As such, the
relativized Hamiltonian (\ref{SH}) has been reduced to that of nonrelativistic
(\ref{Hmn}) formally. As (\ref{Hmn}) indicated, one can view the quantum
average $\langle\nu_{0}\rangle$ as the static energy of the flux-tube (QCD
string) linking the quark $1$ and $2$ \cite{SemaySN:2004,SilvestreSB:JPA2008}.
For more details of the AF method applied to the mesons, see
\cite{KalashnikovaN:PLB2000,SemaySN:2004,SilvestreSB:JPA2008,JIAPH:ijmpa17} .

In the static systems of quark and antiquark, where the total momentum
vanishes ($\mathbf{p}_{1}\mathbf{+p}_{2}=0$), the Hamiltonian (\ref{Hmn})
becomes%
\begin{align}
H(\mu_{1,2},\nu)  &  =\frac{\mathbf{p}^{2}}{2\mu}+\frac{\mu}{2}\left(
\frac{a}{\sqrt{\mu\nu}}\right)  ^{2}r^{2}+\frac{\mu_{m}+\nu}{2}\nonumber\\
&  -\frac{k_{s}(r)}{r}+\frac{m_{1}^{2}}{2\mu_{1}}+\frac{m_{2}^{2}}{2\mu_{2}%
}+V_{0} \label{HCM}%
\end{align}
in which $2\mathbf{p=}$ $\mathbf{p}_{1}\mathbf{-p}_{2}$ defines the relative
momentum $\mathbf{p}$ between quarks, $\mu=\mu_{1}\mu_{2}/\mu_{m}$ is
the reduced effective mass and $\mu_{m}=\mu_{1}+\mu_{2}$.

Given that the AFs $\mu$ and $\nu$ are slow variable and thereby keep constant
effectively, one can diagonalize the first line of (\ref{HCM}), which is
exactly the Hamiltonian of harmonic oscillator. For the whole Hamiltonian
(\ref{HCM}), one can choose the color Coulomb term in the second line of
(\ref{HCM}) as a perturbation. This approximation applies for high excited
states for which the confining force dominates. In the basis of harmonic
oscillator $|nLm\rangle$, the quantized energy $E_{N}(\mu_{1,2},\nu
)=\langle|H(\mu_{1,2},\nu)|\rangle_{nLm}$ of (\ref{HCM}) becomes then%

\begin{align}
E_{N}(\mu_{1,2},\nu)  &  =\frac{a}{\sqrt{\mu\nu}}\left(  N+\frac{3}{2}\right)
+\frac{\mu_{m}+\nu}{2}-\left\langle \frac{k_{s}(r)}{r}\right\rangle
_{N}\nonumber\\
&  +\frac{m_{1}^{2}}{2\mu_{1}}+\frac{m_{2}^{2}}{2\mu_{2}}+V_{0}, \label{Emn}%
\end{align}
where $N=n+L$, with $n$ and $L$ the radial quantum number and the orbital
angular momentum of the bound system, respectively.

For expectation of the color-Coulomb interaction in (\ref{Emn}), we
estimate it by using (\ref{ascut}), giving%

\begin{equation}
\left\langle \frac{k_{s}(r)}{r}\right\rangle \simeq\left\langle \frac
{k_{\infty}}{r+\lambda/\Lambda}\right\rangle =\frac{k_{\infty}}{r^{\ast
}+\lambda/\Lambda}, \label{DEc}%
\end{equation}
where $r^{\ast}$ is some intermediate distance governed by the average size of
the bound system of the quarks. Choosing $r^{\ast}$ to be the expectation
value $\langle|\mathbf{x}_{1}-\mathbf{x}_{2}|\rangle\,$, one has
\begin{equation}
\left\langle \frac{k_{s}(r)}{r}\right\rangle \simeq\frac{k_{\infty}}%
{\langle|\mathbf{x}_{1}-\mathbf{x}_{2}|\rangle+\lambda/\Lambda}. \label{App}%
\end{equation}

In TABLE I(b), the estimations (\ref{DEc}) and (\ref{App}) are checked by
averaging both sides of the equations for $n=0$ ($N=L$). One sees that
(\ref{App}) is valid, more accurately, when $L$ is larger.\medskip

\renewcommand\tabcolsep{0.1cm}
\renewcommand{\arraystretch}{1.8}
\renewcommand{\raggedright}{\leftskip=0pt \rightskip=0pt plus 0cm}
\begin{table}[!htbp]
\flushleft
{{\small TABLE I(b): The color Coulomb term averaged with the harmonic
oscillator }$\Phi=R_{nl}(r)$ {\small is compared to the estimations (\ref{DEc})
and (\ref{App}) for }$L$ {\small from }$0$ {\small to }$7${\small , with
}$\Lambda=0.301$ {\small and }$\lambda=0.1${\small . The harmonic oscillator
length }$a_{H}=0.18${\small , and the unit of the averages is GeV.}}%

\begin{tabular}
[c]{ccccccccc}\hline\hline
$L$ & $0$ & $1$ & $2$ & $3$ & $4$ & $5$ & $6$ & $7$\\\hline
$\left\langle k_{s}(r)/r\right\rangle $ & $0.342$ & $0.247$ & $0.202$ &
$0.174$ & $0.155$ & $0.141$ & $0.131$ & $0.122$\\
$\left\langle \frac{k_{\infty}}{r+\lambda/\Lambda}\right\rangle $ & $0.374$ &
$0.252$ & $0.202$ & $0.174$ & $0.155$ & $0.141$ & $0.130$ & $0.121$\\
$\frac{k_{s}(\infty)}{\langle|\mathbf{r}|\rangle+\lambda/\Lambda}$ & $0.297$ &
$0.223$ & $0.187$ & $0.163$ & $0.147$ & $0.135$ & $0.125$ & $0.118$%
\\\hline\hline
\end{tabular}
\medskip
\end{table}

With the help of the relation $\langle|\mathbf{x}_{1}-\mathbf{x}_{2}%
|\rangle=\nu/a$ in (\ref{min}), Eq. (\ref{Emn}) becomes
\begin{align}
E_{N}(\mu_{1,2},\nu)  &  =\frac{a}{\sqrt{\mu\nu}}\left(  N+\frac{3}{2}\right)
+\frac{\mu_{m}+\nu}{2}\nonumber\\
&  -\frac{k_{\infty}a}{\nu+a_{L}}+\frac{m_{1}^{2}}{2\mu_{1}}+\frac{m_{2}^{2}%
}{2\mu_{2}}+V_{0}. \label{Enu}%
\end{align}
where $a_{L}\equiv\lambda a/\Lambda$. This is the quantized energy of
(\ref{Hmn}) in the enlarged Hilbert space parameterized by the auxiliary
fields and it will give, according to the AF method, the mass spectrum of the
quark-antiquark system considered, provided that $E_{N}(\mu_{1,2},\nu)$ is
minimized in the space of the auxiliary fields.

In Eq. (\ref{Emn}), we write the band quantum number of the harmonic oscillator in
the form $N=n+L$, instead of $N=2n+L$. This is so because when the color
Coulomb term ignored a superfluous ($SU(3)$) dynamical symmetry enters in the
reformulated Hamiltonian (\ref{HCM}) which is originally absent in the
Hamiltonian (\ref{SH}) before the AF method applied: $r\rightarrow r^{2}$.
Such a $SU(3)$ symmetry, known to exist in three-dimensional isotropic
harmonic oscillator (see \cite{Elliott:LondonA58,Schiff68}), brings some
unphysical "accidental" degeneracy in the radially motion and should be removed.

One simple way to remove the above unphysical symmetry is to go back to the
Hamiltonian (\ref{SH}) to consider a one dimensional problem of a massless quark
$1$ moving in the force field $a|x|$ along the radial direction, with $x=r/2$
the radial coordinate of the quark $1$ in the CM system. In this case, the
dynamics simplifies
\begin{equation}
H_{c}\equiv\frac{1}{2}H_{0}=\sqrt{p_{x}^{2}+\left(  \frac{L_{q}}{x}\right)
^{2}}+a|x|. \label{Hpr}%
\end{equation}
When $L_{q}=0\,$ the WKB quantization condition for Eq. (\ref{Hpr}) gives
\begin{equation}
(n+b)\pi=\int_{x_{-}}^{x_{+}}p_{x}dx=\int_{x_{-}}^{x_{+}}dx[M_{c}-a|x|].
\label{WKB}%
\end{equation}
Here, $x_{\pm}=\pm M_{c}/a$ are two classical turning points given by the
condition $M_{c}=a|x_{\pm}|$, the constant $b$ depends on the boundary
conditions, and $2M_{c}=M_{n}$ is the mass of the quark-antiquark system. Up
integration of (\ref{WKB}), one has $(n+b)\pi=M_{c}^{2}/a$. Same
analysis applies for quark $2$ ($x=-r/2$) so that one can find a quantization condition for the
whole quark-antiquark system, only that the range of quark motion need to be halved
since $x\rightarrow-x$ reflection makes no difference to the meson spectrum.
One finds then, by mapping $n\rightarrow n/2$,
\begin{equation}
M_{n}^{2}=2\pi a(n+2b). \label{Mn2}%
\end{equation}
This confirms the linear relation $M^{2}\sim$ $n+L$ claimed in (\ref{MJ}) by
simply comparing (\ref{Mn2}) with the well-known linear relation $M_{J}%
^{2}\propto2\pi aL$ that is derived from the rotating string
picture \cite{Nambu:D74,Goto:71}.

The relation $M^{2}\propto$ $n+L$ has also been suggested by Afonin et.
al. \cite{Afonin:BA06,Afonin:07} and Bicudo \cite{Bicudo:D07}. Experimental
evidences in favor of this relation were given in \cite{AnisovichAS:PRD00,Afonin:07}. We will
show, in the following section, that the formal discrepancy of the
harmonic-oscillator-like energy (\ref{Enu}) with the linear Regge relation
(\ref{MJ}) can be removed by showing $\sqrt{\mu\nu}\sim\sqrt{N}$ in the large
$N$ limit.

\section{Mass formula and quasi-linear Regge trajectories}\label{sec3}

As stated earlier, to solve the model (\ref{SH}) with the AF method, one has
to minimize the energy (\ref{Enu}) in the space of the auxiliary fields. This
amounts to solving simultaneously the three constraints $\partial_{A}E_{N}%
(\mu_{1,2},\nu)=0$ $(A=\mu_{1},\mu_{2},\nu)$, which are explicitly
\begin{align}
\frac{a_{N}\nu}{\sqrt{(\mu\nu)^{3}}}\left(  \frac{\mu_{2}}{\mu_{m}}\right)
^{2}  &  =1-\frac{m_{1}^{2}}{\mu_{1}^{2}},\label{cons1}\\
\frac{a_{N}\nu}{\sqrt{(\mu\nu)^{3}}}\left(  \frac{\mu_{1}}{\mu_{m}}\right)
^{2}  &  =1-\frac{m_{2}^{2}}{\mu_{2}^{2}},\label{cons2}\\
\frac{a_{N}\mu}{\sqrt{(\mu\nu)^{3}}}  &  =1+\frac{2k_{\infty}a}{(\nu
+a_{L})^{2}}, \label{cons3}%
\end{align}
with $k_{\infty}\equiv 4\alpha_{s}(\infty)/3$, and
\begin{equation}
a_{N}\equiv a(N+3/2). \label{aN}%
\end{equation}

For unflavored meson $n\bar{n}$ ($n$ stands for $u$ or $d$ quarks) the
bare quark mass $m_{i}$ should be small, much smaller than the effective mass
$\mu_{i}$. Notice that the average interquark distance $l=\langle r\rangle$ is
about $\nu/a$, one can estimate, for the high excited states ($\nu$ is large),
\begin{equation}
\frac{a}{\nu^{2}}\sim\frac{a}{(la)^{2}}=\left(  \frac{r_{G}}{l}\right)
^{2}\ll1, \label{Av}%
\end{equation}
where $r_{G}\sim\sqrt{1/a}$ is the characteristic size of the ground-state
meson\footnote{The characteristic size of a meson in the ground state can be roughly determined by
the balancing two terms of the potential energy $ar$ and $1/r$. This gives
$r_{G}^{2}\sim1/a$.}.

Given that the bare masses $m_{i}$ $\ll$ $\mu_{i}$, one can solve
Eqs. (\ref{cons1}) and (\ref{cons2}). Up to the leading order of $\Delta
m^{2}/\mu_{m}^{2}$, where $\Delta m^{2}\equiv m_{1}^{2}-m_{2}^{2}$ and
$\mu_{m}$ is the sum of two effective masses, the results are
\begin{equation}
\mu=\frac{\mu_{m}}{4}=\frac{a_{N}^{2}}{\nu^{3}[1+2ak_{\infty}/N_{\nu}]^{2}},
\label{mui}%
\end{equation}
\begin{align}
\mu_{1}  &  =\frac{\mu_{m}}{2}\left(  1+\frac{\Delta m^{2}}{\mu_{m}^{2}%
}\right)  ,\label{mu12}\\
\mu_{2}  &  =\frac{\mu_{m}}{2}\left(  1-\frac{\Delta m^{2}}{\mu_{m}^{2}%
}\right)  .\nonumber
\end{align}
where $N_{\nu}\equiv(\nu+a_{L})^{2}$. Here, we always assume quark $1$ is
heavier than antiquark $2$ if the quark $1$ is strange while the quark $2$ is nonstrange.

Putting the relations (\ref{mui}) and (\ref{mu12}) into (\ref{Enu}), one has%

\begin{align}
E_{N}(\nu)  &  =\left[  \frac{3}{2}+\frac{3ak_{\infty}}{N_{\nu}}+\frac
{2a_{N}^{2}}{\nu^{4}\chi_{N}^{2}}\right]  \nu-\frac{ak_{\infty}}{N_{\nu}%
^{1/2}}\nonumber\\
&  +\frac{\bar{m}^{2}}{2a_{N}^{2}}\nu^{3}\chi_{N}^{2}+V_{0}. \label{ENv}%
\end{align}
in which
\begin{align}
\chi_{N}  &  =1+2\frac{ak_{\infty}}{N_{\nu}},\nonumber\\
\bar{m}^{2}  &  \equiv\frac{1}{2}(m_{1}^{2}+m_{2}^{2}). \label{cham}%
\end{align}
Minimizing of the energy (\ref{ENv}) by the constraint equation,
$\delta_{\nu}E_{N}(\mu_{0},\nu)=0$, yields%
\begin{align}
&  \frac{3}{2}+\frac{3ak_{\infty}}{N_{\nu}}-\frac{6a_{N}^{2}}{\nu^{4}\chi
_{N}^{2}}-\frac{4ak_{\infty}\nu}{N_{\nu}^{3/2}}+\frac{16ak_{\infty}a_{N}^{2}%
}{\nu^{3}\chi_{N}^{3}N_{\nu}^{3/2}}\nonumber\\
&  =\frac{\bar{m}^{2}}{2a_{N}^{2}}\left(  8ak_{\infty}\frac{\nu^{3}\chi_{N}%
}{N_{\nu}^{3/2}}-3\nu^{2}\chi_{N}^{2}\right)  , \label{AF2}%
\end{align}

Eq. (\ref{AF2}) is nonlinear and quite involved for analytical treatment. What
is more involved here is that the knowledge of the interquark
interaction (\ref{V}) is not complete. Bearing in mind of this limitation in
the interquark interaction (\ref{V}), we firstly solve (\ref{AF2}) in the
large $N$ limit using the nonperturbative method of homotopic analysis (HA)
\cite{Liao}, and then extend the Regge-like solution obtained thereby to the
low-$N$ case by quasi-linearizing the ensuing mass formula, up to the leading order of $1/N$.

In the large $N$ limit, we assume $\nu^{2}\sim a_{N}\gg1$, which can be shown
by solving (\ref{AF2}) numerically (see FIG. 2 and Table II). Taking
$a/\nu^{2}\rightarrow0$, Eq. (\ref{AF2}) simplifies%
\begin{equation}
\frac{3}{2}-\frac{6a_{N}^{2}}{\nu^{4}}=\frac{\bar{m}^{2}}{2a_{N}^{2}}\left(
8ak_{\infty}-3\nu^{2}\right)  , \label{Af2}%
\end{equation}
where $N_{\nu}\rightarrow\nu^{2}$ and $\chi_{N}\rightarrow1$ have been
applied. It follows from (\ref{Af2}), by treating the mass term as a perturbation,
that
\begin{equation}
\nu^{2}=2a_{N}-2\bar{m}^{2}=2a\left(  N+\frac{3}{2}\right)  -m_{1}^{2}%
-m_{2}^{2}, \label{vva}%
\end{equation}
which agrees qualitatively, in the massless limit ($m_{1}\sim m_{2}=0$), with
the Regge phenomenology: $\nu^{2}\propto N=n+L$.

Given the solution (\ref{vva}), one can use the method of HA \cite{Liao} to
solve Eq. (\ref{AF2}). The result is (Appendix A)
\begin{equation}
\nu_{N}^{2}=2a_{N}+ak_{\infty}\left(  h-2+\frac{4a_{L}}{\sqrt{2a_{N}}}\right)
-2\bar{m}^{2}\left[  1-\frac{4e_{N}}{3}\right]  , \label{vsolu}%
\end{equation}
in which
\begin{equation}
e_{N}\equiv\frac{ak_{\infty}}{a_{N}}=\frac{2k_{\infty}}{2N+3}. \label{eJ}%
\end{equation}
Here, $h$ is the accelerating factor \cite{Liao} remained to be fixed
empirically. We fix $h$ simply by comparing $\nu^{2}$ in (\ref{vsolu}) with
the numerical solution to (\ref{AF2}). The results are shown in FIG. 2 and
\textrm{Table II(a)}. One sees, quite remarkably, that the solution $\nu^{2}$ to
Eq. (\ref{AF2}) rises almost linearly with $L$, both for that of analytical (solid
line) and of numerical (dots in FIG. 2).
%TCIMACRO{\FRAME{ftbpFU}{3.4714in}{2.6567in}{0pt}{\Qcb{The analytical solution
%(solid line) (\ref{vsolu}) to the equation (\ref{AF2}) compared to the
%numerical solution for the $L$-dependence of $\nu^{2}$. }}{}{nu2-l.eps}%
%width 3.4714in;  height 2.6567in;  depth 0pt;  original-width 9.0278in;
%original-height 6.8952in;  cropleft "0";  croptop "1";  cropright "1";
%cropbottom "0";  filename 'Nu2-L.eps';file-properties "XNPEU";}} }%
%BeginExpansion
\begin{figure}
[ptb]
\begin{center}
\includegraphics[
width=3.4in
]%
{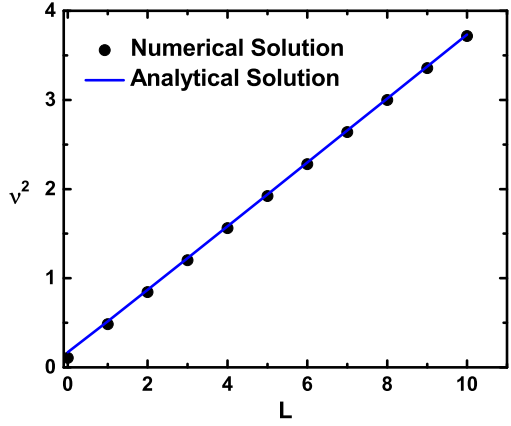}%
\caption{The analytical solution (solid line) (\ref{vsolu}) to the equation
(\ref{AF2}) compared to the numerical solution for the $L$-dependence of
$\nu^{2}$. }%
\end{center}
\end{figure}
%EndExpansion

\renewcommand\tabcolsep{0.13cm}
\renewcommand{\arraystretch}{1.8}
\renewcommand{\raggedright}{\leftskip=0pt \rightskip=0pt plus 0cm}
\begin{table}[!htbp]
\flushleft
{{\small TABLE II(a): The analytical (solid line) solutions to (\ref{AF2}) and numerical (dots)
solutions to the auxiliary field equations (\ref{cons1}) through (\ref{cons3})
for \{}$\nu^{2}${\small \}. The deviations between two solutions are also listed in the fourth row. The accelerating factor }$h=-0.88${\small .}}%

\begin{tabular}
[c]{c|ccccccc}\hline\hline
$L$ & $0$ & $1$ & $2$ & $3$ & $4$ & $5$ & $6$\\\hline
$%
\begin{array}
[c]{r}%
\text{Numerical}\\
\nu^{2}\text{[GeV}^{2}\text{]}%
\end{array}
$ & ${\small 0.107}$ & ${\small 0.483}$ & ${\small 0.844}$ & ${\small 1.203}$
& ${\small 1.562}$ & ${\small 1.921}$ & ${\small 2.280}$\\\hline
$%
\begin{array}
[c]{r}%
\text{Analytical}\\
\nu^{2}\text{[GeV}^{2}\text{]}%
\end{array}
$ & ${\small 0.169}$ & ${\small 0.516}$ & ${\small 0.869}$ & ${\small 1.225}$
& ${\small 1.582}$ & ${\small 1.940}$ & ${\small 2.298}$\\\hline
Ana.$-$Num. & ${\small 0.063}$ & ${\small 0.033}$ & ${\small 0.025}$ &
${\small 0.022}$ & ${\small 0.020}$ & ${\small 0.019}$ & ${\small 0.018}%
$\\\hline\hline
\end{tabular}
\medskip\medskip
\end{table}

\renewcommand\tabcolsep{0.05cm}
\renewcommand{\arraystretch}{1.8}
\renewcommand{\raggedright}{\leftskip=0pt \rightskip=0pt plus 0cm}
\begin{table}[!htbp]
\flushleft
{{\small TABLE II(b): The effective masses \{}$\mu_{1},\mu_{2}${\small \}
solved numerically from (\ref{cons1}) through (\ref{cons2}), compared to their
analytical values given by (\ref{mu12}).}}%

\begin{tabular}
[c]{c|ccccccc}\hline\hline
${\small L}$ & ${\small 0}$ & ${\small 1}$ & ${\small 2}$ & ${\small 3}$ &
${\small 4}$ & ${\small 5}$ & ${\small 6}$\\\hline
$\text{Num. }%
\begin{array}
[c]{r}%
\mu_{1}\\
\mu_{2}%
\end{array}
$ & $%
\begin{array}
[c]{r}%
{\small 0.483}\\
{\small 0.492}%
\end{array}
$ & $%
\begin{array}
[c]{r}%
{\small 0.527}\\
{\small 0.536}%
\end{array}
$ & $%
\begin{array}
[c]{r}%
{\small 0.601}\\
{\small 0.609}%
\end{array}
$ & $%
\begin{array}
[c]{r}%
{\small 0.670}\\
{\small 0.676}%
\end{array}
$ & $%
\begin{array}
[c]{r}%
{\small 0.733}\\
{\small 0.739}%
\end{array}
$ & $%
\begin{array}
[c]{r}%
{\small 0.791}\\
{\small 0.797}%
\end{array}
$ & $%
\begin{array}
[c]{r}%
{\small 0.846}\\
{\small 0.851}%
\end{array}
$\\\hline
$\text{Ana. }%
\begin{array}
[c]{r}%
\mu_{1}\\
\mu_{2}%
\end{array}
$ & $%
\begin{array}
[c]{r}%
{\small 0.402}\\
{\small 0.391}%
\end{array}
$ & $%
\begin{array}
[c]{r}%
{\small 0.506}\\
{\small 0.497}%
\end{array}
$ & $%
\begin{array}
[c]{r}%
{\small 0.590}\\
{\small 0.582}%
\end{array}
$ & $%
\begin{array}
[c]{r}%
{\small 0.663}\\
{\small 0.656}%
\end{array}
$ & $%
\begin{array}
[c]{r}%
{\small 0.728}\\
{\small 0.721}%
\end{array}
$ & $%
\begin{array}
[c]{r}%
{\small 0.787}\\
{\small 0.781}%
\end{array}
$ & $%
\begin{array}
[c]{r}%
{\small 0.842}\\
{\small 0.837}%
\end{array}
$\\\hline\hline
\end{tabular}
\medskip
\end{table}

Hence, Eqs. (\ref{cons1}) through (\ref{cons3}) are solved by (\ref{mu12}) and
(\ref{vsolu}). Putting them into (\ref{ENv}) yields
\begin{equation}
M_{\bar{q}q}=w_{N}\nu_{N}+\frac{A_{N}}{\nu_{N}}+V_{0} \label{Mqq}%
\end{equation}
in which $\nu_{N}$ is given by (\ref{vsolu}) and
\begin{align}
w_{N}  &  =\frac{3}{2}+\frac{3ak_{\infty}}{N_{\nu}}+\frac{2a_{N}^{2}}{\nu
^{4}\chi_{N}^{2}}\label{wN}\\
A_{N}  &  =2\bar{m}^{2}\chi_{N}^{2}\left(  \frac{\nu_{N}^{2}}{2a_{N}}\right)
^{2}-\frac{ak_{\infty}}{1+a_{L}/\nu_{N}} \label{AN}%
\end{align}

Since $\nu_{N}$ is solved from (\ref{cons1}) through (\ref{cons3}) in the
relatively large-$N$ region, and the approximation (\ref{App}) for the color
Coulomb interaction applies better in the large distance regime, the
prediction (\ref{Mqq}), obtained by the quark model (\ref{SH}) combined with
AF method, should be more reliable for the high excited $\bar{q}q$ mesons.
When $N$ is very large, $N_{\nu}$ as well as $\nu_{N}^{2}\rightarrow2a\left(
N+3/2\right)  $, $\chi_{N}\rightarrow1$ and $w_{N}\rightarrow2$, which leads, by
Eq. (\ref{Mqq}), to
\begin{equation}
M_{\bar{q}q}=2\nu_{N}+\left(  2\bar{m}^{2}-ak_{\infty}\right)  /\nu_{N}%
+V_{0}\,\text{,when }N\gg1, \label{MqqN}%
\end{equation}
or
\[
(M_{\bar{q}q}-V_{0})^{2}\propto8aN,\text{when }N\gg1,
\]
This corresponds to the slope $(8a)^{-1}$ for the linear Regge trajectory on
the $(N,(M_{\bar{q}q}-V_{0})^{2})$ plot that is predicted by the relativized
quark model \cite{KangS:75,VeseliO:B96}. It is to be compared with the slope
$1/(2\pi a)$ predicted by the relativistic string model
\cite{Goto:71,Nambu:D74}.

We remark that when $N$ is large Regge linearity stems from the first and
second term in (\ref{Enu}), which appears to be of the harmonic-oscillator
form: $a(N+3/2)+const$, as the most quark models with harmonic-oscillator-like
confinement predicted. This changes, however, in our model due to the
constraints (\ref{cons1})-(\ref{cons2}) of the AF fields. The solutions
(\ref{mui}) and (\ref{mu12}) indicate $\mu=\mu_{m}/4\sim\sqrt{N}$ and $\nu
\sim\sqrt{N}$ (namely,$\sqrt{\mu\nu}\sim\sqrt{N}$) in the large $N$ limit.
Thus, when $N$ is large the first and second terms in (\ref{Enu}) scale as
$2\sqrt{2a_{N}}\sim\sqrt{N}$, hence the linear Regge behavior: $E_{N}%
^{2}\propto N$.

The mass relation (\ref{Mqq}) is inadequate in the low-$N$ region for two
reasons rigorously. The first is obviously that Eqs. (\ref{vsolu}), (\ref{mui}) and
(\ref{mu12}) are not valid in the low-$N$ region, as seen in Table II (a,b),
and that the approximation (\ref{App}) does not apply in the low-lying states,
as roughly shown in Table II (a,b). The second, more serious, is that the short
distance behavior of the interquark interaction is far from established
\cite{Bali01,KawanaiS:12}, e.g., the running of strong coupling remains
unclear \cite{BrodskyT:B04,AguilarMN:A04,ShirkovSol:97,Ganbold:D10}. Thus, to
find the mass relation for the low-$N$ region, we resort to the approximate
linearity of the Regge trajectories that is established experimentally in
meson spectrums \cite{AnisovichAS:PRD00} to constraint the prediction
(\ref{Mqq}). By squaring (\ref{Mqq}), it follows that
\begin{align}
\left(  M_{\bar{q}q}-V_{0}\right)  ^{2}  &  =2aw_{N}^{2}\left[  N+\frac{3}%
{2}-\frac{\bar{m}^{2}}{a}\left(  1-\frac{4e_{N}}{3}\right)  \right.
\nonumber\\
&  +\left.  \frac{k_{\infty}}{2}\left(  h-2+\frac{4a_{L}}{\sqrt{2a_{N}}%
}\right)  +D_{N}\right]  , \label{QRegg}%
\end{align}
where
\begin{equation}
D_{N}=\frac{A_{N}}{aw_{N}}+\frac{a}{2\nu_{N}^{2}}\left(  \frac{A_{N}}{aw_{N}%
}\right)  ^{2}. \label{wAN}%
\end{equation}
The $N$-dependence of $\left(  M_{\bar{q}q}-V_{0}\right)  ^{2}$ in
(\ref{QRegg}) is nonlinear formally when compared to the Chew-Frautschi
plot \cite{ChewF:61}. The constraining of (\ref{Mqq}) and extrapolating it to
the relatively low-$N$ region can be done by making (\ref{QRegg}) quasi-linear
in $N$. We note firstly that (\ref{QRegg}) is comparable to the linear Regge
trajectories (\ref{MJ}), provided that the $V_{0}$ is small when compared to
the meson scale, $V_{0}/M_{\bar{q}q}\ll1$. If we rewrite (\ref{QRegg}) in the
form
\begin{equation}
\alpha_{N}^{\prime}\left(  M_{\bar{q}q}-V_{0}\right)  ^{2}=N-\alpha_{N}(0),
\label{QRe}%
\end{equation}
in which the trajectory parameters $\alpha_{N}^{\prime}$ and $\alpha_{N}(0)$
are,
\begin{equation}
\frac{1}{\alpha_{N}^{\prime}}=2aw_{N}^{2}=\frac{9a}{2}\left[  1+\frac{\left(
2a_{N}/\nu_{N}^{2}\right)  ^{2}}{3\chi_{N}^{2}}+\frac{2ak_{\infty}}{N_{\nu}%
}\right]  ^{2}, \label{Slope}%
\end{equation}%
\begin{equation}
-\alpha_{N}(0)=\frac{3}{2}-\frac{\bar{m}^{2}}{a}\left(  1-\frac{4e_{N}}%
{3}\right)  +\frac{k_{\infty}}{2}\left(  h-2+\frac{4a_{L}}{\sqrt{2a_{N}}%
}\right)  +D_{N}, \label{a12}%
\end{equation}
respectively, then one can quasi-linearize (\ref{QRegg}) by expanding (\ref{Slope}) and
(\ref{a12}) on $1/N$. To order of $1/N$, the last term
$D_{N}$ in (\ref{a12}) becomes
\begin{align}
D_{N}  &  \simeq\frac{\bar{m}^{2}}{a}\left(  1-\frac{9\bar{m}^{2}}{4a_{N}%
}\right)  +\frac{k_{\infty}}{2}\left(  \frac{a_{L}}{\sqrt{2a_{N}}}-1\right)
\label{Da}\\
&  +\frac{k_{\infty}}{16a_{N}}\left(  ak_{\infty}(7-2h)-4a_{L}^{2}+4\bar
{m}^{2}(5h-3)\right)  ,
\end{align}
which leads to, when putting to (\ref{a12}),
\begin{equation}
-\alpha_{N}(0)=\frac{3}{2}+\frac{k_{\infty}}{2}\left(  h-3+\frac{5a_{L}}%
{\sqrt{2a_{N}}}\right)  +\frac{k_{\infty}\bar{m}^{2}}{4a_{N}}(5h-3).
\label{int}%
\end{equation}
The similar relation for the inverse slope (\ref{Slope}) is%
\begin{equation}
\frac{1}{\alpha_{N}^{\prime}}=8a\left[  1+\frac{1}{2}\frac{3k_{\infty}}%
{2N+3}\left(  1-\frac{h}{3}-\frac{2a_{L}}{\sqrt{2a_{N}}}\right)  +\frac
{\bar{m}^{2}}{2a_{N}}\right]  ^{2}. \label{ina}%
\end{equation}
When $N$ is very large Eq. (\ref{ina}) tends to a inverse slope: $\lim
_{N\rightarrow\infty}$($\alpha_{N}^{\prime-1})=8a$, in consistent with claims
in Refs. \cite{KangS:75,VeseliO:B96}.

One sees from (\ref{int}) and (\ref{ina}) that the slope depends upon the
dimensional parameters $a$ and upon the dimensionless parameters $k_{\infty}$,
$\sqrt{a}/\Lambda$ and $\bar{m}^{2}/a$ weakly (suppressed by $N$ or $\sqrt{N}%
$) while the intercept depends upon $k_{\infty}\,$ strongly and also upon $\sqrt
{a}/\Lambda$ and $\bar{m}^{2}/a$ weakly. Given (\ref{int}) and (\ref{ina}),
one rewrite (\ref{QRe}) in the front of analytical mass formula for light mesons,
\begin{equation}
M_{q\bar{q}}=\left(  1+K_{N}\right)  \sqrt{8a\left[  N-\alpha_{N}(0)\right]
}+V_{0}, \label{MMq}%
\end{equation}
where
\begin{equation}
K_{N}=\frac{3ak_{\infty}}{4a_{N}}\left(  1-\frac{h}{3}-\frac{2\lambda\sqrt{a}%
}{\Lambda\sqrt{2N+3}}\right)  +\frac{m_{1}^{2}+m_{2}^{2}}{2a(2N+3)},
\label{lN}%
\end{equation}%
\begin{equation}
-\alpha_{N}(0)=\frac{3}{2}+\frac{k_{\infty}}{2}\left(  h-3+\frac{5\lambda
\sqrt{a}}{\Lambda\sqrt{2N+3}}\right)  +\frac{k_{\infty}\bar{m}^{2}}{2a}%
\frac{5h-3}{2N+3}. \label{al0}%
\end{equation}
The formula (\ref{MMq}) is the main result in this work. We see that the
flavor dependence enters explicitly through the mass term $(m_{1}^{2}%
+m_{2}^{2})/a$. The following remarks are in order:

(i) The Hamiltonian (\ref{SH}), $H=\mu_{1}+\mu_{2}+V$, becomes almost
independent of the quark masses in the light-light limit $m_{1,2}\rightarrow0$
for which $\mu_{i}=\sqrt{|\mathbf{p}|^{2}+m_{i}^{2}}\rightarrow|\mathbf{p}|$.
The same it true when $L$ is large since $|\mathbf{p}|^{2}$ has the
expectation $\sim N$ in the harmonic basis $|nL\rangle$. See (\ref{mui}) and
(\ref{mu12}). The $m_{1,2}$-dependence of the system mass is thereby
suppressed by $N$. This accounts for the asymptotic flavor independence
happened in Table II(b) that $\mu_{1,2}$ tend to be same with $L$ increases.

(ii) In spite of assumption $a_{N}\propto N\gg1$ in obtaining (\ref{MMq}) and
(\ref{Mqq}) from the Hamiltonian (\ref{SH}), the quasi-linearizing of the model
prediction (\ref{QRegg}) makes it applicable in the low-excited states, thanks
to the Regge phenomenology for light mesons.

(iii) The mass formula (\ref{MMq}) goes beyond the native prediction of the
relativized quark model in that it employs merely the large-$N$ asymptotic behaviors of the model
spectrum that is implied in relativistic quark model.

(iv) The ultra-relativistic contributions from QCD string rotating to the
orbital angular momentum and to the energy of meson has not taken into account in
(\ref{MMq}), which can otherwise enhance the confining parameter $a$ by a
factor of $8a/(2\pi a)=4/\pi$ in the high excited states of mesons, which will
be discussed in the section 4 and section 5.

\section{Numerical results and discussions}\label{sec4}

\renewcommand\tabcolsep{0.7cm}
\renewcommand{\arraystretch}{1.8}
\renewcommand{\raggedright}{\leftskip=0pt \rightskip=0pt plus 0cm}
\begin{table*}[!htbp]
\flushleft
{{\small TABLE III: The members in six families and their linear fit to the
observed masses squared in \cite{Patrignani:C16}. The squared bracket is used
to indicate that }${\small \eta/h}$ {\small family has an abnormal slope in
nonstrange sector. The mark EF in the last row for the meson }$f_{4}%
^{^{\prime}}(4^{++})$ {\small indicates that the data comes from the quark
model prediction in Ref. \cite{EbertFG:D09}.}}%

\begin{tabular}
[c]{cccccc}\hline\hline
{\small Traj.} & {\small \ mesons (}$J^{PC}${\small )}$\ \ $ & $%
\begin{array}
[c]{r}%
\text{{\small Linear fit}}\\
M^{2}\text{(GeV}^{2}\text{)}%
\end{array}
$ & $%
\begin{array}
[c]{c}%
\text{{\small Slope}}\\
\multicolumn{1}{r}{{\small \alpha}^{\prime}\text{[GeV}^{-2}\text{]}}%
\end{array}
$ & {\small Intercept} & $%
\begin{array}
[c]{c}%
1/{\small (2\pi\alpha^{\prime})}\\
\multicolumn{1}{r}{\text{GeV}^{2}}%
\end{array}
$\\\hline
${\small \pi/b}$ & $\left\{
\begin{array}
[c]{r}%
b_{1}(1^{+-}),\pi_{2}(2^{-+}),\\
b_{3}(3^{+-}),\pi_{4}(4^{-+})
\end{array}
\right.  $ & ${\small 0.3770+1.199L}$ & ${\small \allowbreak0.834}$ &
${\small -0.314}$ & ${\small 0.191}$\\
${\small \rho/a}$ & $\left\{
\begin{array}
[c]{r}%
\rho(1^{--}),a_{2}(2^{++}),\\
\rho_{3}(3^{--}),a_{4}(4^{++}),\\
\rho_{5}(5^{--}),a_{6}(6^{++})
\end{array}
\right.  $ & ${\small 0.6328+1.120L}$ & ${\small \allowbreak0.893}$ &
${\small -0.565}$ & ${\small 0.178}$\\
\lbrack${\small \eta/h}$] & $\left\{
\begin{array}
[c]{c}%
\eta(0^{-+}),h_{1}(1^{+-}),\\
\multicolumn{1}{r}{\eta_{2}(2^{-+}),h_{3}(3^{+-}),}\\
\eta_{4}(4^{-+})
\end{array}
\right.  $ & ${\small 0.1668+1.297L}$ & ${\small \allowbreak0.771\,}$ &
${\small -0.128}$ & ${\small 0.206}$\\
${\small \omega/f}$ & $\left\{
\begin{array}
[c]{r}%
\omega(1^{--}),f_{2}(2^{++}),\\
\omega_{3}(3^{--}),f_{4}(4^{++}),\\
\omega_{5}(5^{--}),f_{6}(6^{++})
\end{array}
\right.  $ & ${\small 0.5877+1.115L}$ & ${\small \allowbreak0.897}$ &
${\small -0.527}$ & ${\small 0.177}$\\
${\small K}^{\ast}$ & $\left\{
\begin{array}
[c]{c}%
K^{\ast}(1^{-}),K_{2}^{\ast}(2^{+}),\\
\multicolumn{1}{r}{K_{3}^{\ast}(3^{-}),K_{4}^{\ast}(4^{+}),}\\
K_{5}^{\ast}(5^{-})
\end{array}
\right.  $ & ${\small 0.7868+1.191L}$ & ${\small \allowbreak0.840}$ &
${\small -0.661}$ & ${\small 0.189}$\\
${\small \phi/f}^{\prime}$ & $\left\{
\begin{array}
[c]{c}%
\phi(1^{--}),f_{2}^{\prime}(2^{++}),\\
\multicolumn{1}{r}{\phi_{3}(3^{--}),f_{4}^{^{\prime}}(4^{++})^{\text{EF}}}%
\end{array}
\right.  $ & ${\small 0.9838+1.325L}$ & ${\small \allowbreak0.755}$ &
${\small -\allowbreak0.742}$ & ${\small 0.211}$\\\hline\hline
\end{tabular}
\medskip
\end{table*}

In this section, we confront the mass formula (\ref{MMq}) with the experiments
and other approaches. As explained in the text, we are of mainly interest in
the orbitally excited states, where our approximation is expected to work
best. For this, we choose six families of light mesons, marked by $\pi/b$,
$\rho/a$, $\eta/h$, $\omega/f$, $K^{\ast}$ and $\phi/f^{\prime}$. The members
we pick for the trajectories are always the lightest known states with
appropriate quantum numbers. In each family, the quantum numbers of $P$ and
$C$ alternate their values across the trajectory provided that they are mainly
made of $\bar{q}q$ system with the parity $P=(-)^{L+1}$ and $C=(-)^{L+S}$.
Furthermore, the $J=0$ state in the $\pi/b$-trajectory, which is actually the
pion, has been excluded due to its abnormally low mass.

In Table III, the selected family members are shown explicitly, together with
the linear fit for the observed mass squared $M^{2}$ v.s. $L$, with the data
taken entirely from the Particle Data Group's (PDG) 2016 Review of Particle
Physics \cite{Patrignani:C16}. We also list the corresponding slope
$\alpha^{\prime}$, intercept and the MS error for linear fit defined by
$\chi_{MS}^{2}=\sum_{L}(M_{L}^{Th}-M_{L}^{Exp})^{2}/L_{\max}$ where the index
$L$ runs from $0$ ($1$ for the $\pi/b$ trajectory) to the maximal value
$L_{\max}$ of the orbital angular momentum. From the linear fit shown in Table
III one sees that the linear relation (\ref{MJ}) applies for the trajectories
of $\rho/a$ and $\omega/f$ for which intercept is about $-0.5$, but is
violated for the $\pi/b$, $\eta/h$ and $\phi/f^{\prime}$ trajectories for
which the intercepts are about $-0.3$,$-0.1$ and $-0.74$, respectively. The
$K^{\ast}$ trajectory can fit the relation (\ref{MJ}) very roughly.

We use the mass formula (\ref{MMq}), with $K_{N}$ and $-\alpha_{N}(0)$ given
by (\ref{lN}) and (\ref{al0}), to map the observed masses in each family in
Table III, guided globally by corresponding linear fit in the Table III. For
the family of $\eta/h$, with the ideal mixing $\Phi(\eta/h)=\left(  \sqrt
{2}nn-2\bar{s}s\right)  /\sqrt{6}$ assumed for the flavor content, we use the
mass formula
\begin{equation}
M_{\eta/h}=\frac{1}{3}M_{n\bar{n}}+\frac{2}{3}M_{s\bar{s}}, \label{Metah}%
\end{equation}
with the masses $M_{n\bar{n}}$ and $M_{s\bar{s}}$ given by (\ref{MMq}) and
$2\bar{m}^{2}$($=m_{1}^{2}+m_{2}^{2}$)$=2m_{ud}^{2}$ and $2m_{s}^{2}$,
respectively. The results for the optimal parameters ($a,\Lambda$,$V_{0}$)
determined are shown in Table IV. Due to its abnormal nature, we always take the family of
$\eta/h$ to be abnormal in this work. \medskip

\renewcommand\tabcolsep{0.4cm}
\renewcommand{\arraystretch}{1.8}
\renewcommand{\raggedright}{\leftskip=0pt \rightskip=0pt plus 0cm}
\begin{table}[!htbp]
\flushleft
{{\small TABLE IV: The parameters in the mass formula (\ref{MMq}) mapping the
experimental spectrum of the meson families considered, including the
confining parameter (}$a${\small ), the low-energy cutoff (}$\Lambda$)
{\small averaged and the vacuum constant }$V_{0}${\small . The squared bracket
indicates that }${\small \eta/h}$ {\small family has an abnormal slope in
nonstrange sector.}}%

\begin{tabular}
[c]{ccccc}\hline\hline
Traj. & $h$ & $a$(GeV$^{2}$) & $\Lambda$(GeV) & $V_{0}$(GeV)\\\hline
$\pi/b$ & $-0.88$ & $0.153$ & $0.303$ & $-0.237$\\
$\rho/a$ & $-0.48$ & $0.145$ & $0.152$ & $-0.200$\\
\lbrack$\eta/h$] & $-0.48$ & $0.178$ & $0.239$ & $-0.431$\\
$\omega/f$ & $-0.28$ & $0.146$ & $0.133$ & $-0.245$\\
$K^{\ast}$ & $-0.08$ & $0.140$ & $0.203$ & $-0.103$\\
$\phi/f^{\prime}$ & $-0.18$ & $0.152$ & $0.131$ & $-0.073$\\\hline\hline
\end{tabular}
$\ \ \ \ $ \vspace{3mm}
\end{table}

As can seen in the Table IV, the values of $V_{0}$ are all negative, as argued
in \cite{Gromes:Z81} for the non-relativistic limit of the Bethe-Salpeter
equation. Furthermore, apart from the family of $\eta/h$ the model parameters
($a,\Lambda$) are taken values around
\begin{equation}%
\begin{array}
[c]{r}%
a=0.147\pm0.005\text{GeV}^{2}\text{,}\\
\Lambda=0.184\pm0.072\text{GeV.}%
\end{array}
\label{aglo}%
\end{equation}
The small fluctuation implies, especially for $a$, that $a$ and
$\Lambda$ keep the same approximately. In contrast, the linear fit in Table
III (the last column) implies an averaged string tension (assuming the QCD string
picture)
\begin{equation}
a\text{(Linear fit)}=0.189\pm0.0135\text{GeV}^{2}\text{,} \label{LinF}%
\end{equation}
with much larger fluctuation than that in (\ref{aglo}). We note that in the
above analysis $\eta/h$ trajectory is excluded as an exceptional case. In this
sense, the confining parameter $a$ is almost universal. A detailed comparison
of the result (\ref{aglo}) with other predictions in the literatures is shown
in Table V. \medskip

\renewcommand\tabcolsep{0.2cm}
\renewcommand{\arraystretch}{1.8}
\renewcommand{\raggedright}{\leftskip=0pt \rightskip=0pt plus 0cm}
\begin{table}[!htbp]
\flushleft
{{\small TABLE V: Comparison of predictions for the confining parameter }%
$a${\small . Here, QM stands for quark model, Rel. for relativistic, SSE for
Spinless Salpeter equation. LGT stands for lattice QCD.}}%

\begin{tabular}
[c]{ccc}\hline\hline
Reference & $a$ (GeV$^{2}$) & Method\\\hline
Present work & $0.147$ & Rel. QM + AFM + quasi-linearizing\\
\cite{VeseliO:B96} & $0.142$ & Rel. QM + Bjorken Sum rule\\
\cite{KangS:75} & $0.30$ & Rel. QM + Scaling\\
\cite{Durand:D84} & $0.180$ & Semi-relativistic QM\\
{\small \cite{GodfreyIsgur:85}} & $0.18$ & Relativized QM\\
\cite{Jacobs:D86} & $0.192$ & Semi-relativistic QM\\
\cite{LuchaRS:D92} & $0.211$ & SSE + Smooth transition potential\\
\cite{FulcherCY:D93} & $0.191$ & Rel. Kinematics + Linear + Coulomb\\
\cite{HwangK:D96} & $0.183$ & Semi-relativistic QM\\
{\small \cite{EbertFG:D09}} & $0.24$ & Rel. pseudopotential QM\\
{\small \cite{SonnenscheinW:JHEP14}} & $0.18$ & Massive Rel. string\\
\cite{KawanaiS:12} & $0.155(19)$ & Unquenched LGT\\\hline\hline
\end{tabular}
\medskip
\end{table}

One can observe from Table V that the values of $a$ is close to that predicted
by Veseli and Olsson \cite{VeseliO:B96} and the linearly confining parameter
for the heavy-quarkonia in lattice QCD \cite{KawanaiS:12}, whereas it is
smaller, about $20\%$, than that in the other quark models cited. In the end
of Section 4 and in the Section 5, we address this issue in details.

We list the mass predictions by the formula (\ref{MMq}) and the experimentally
observed masses in Table VI. As seen there, an good agreement is achieved
between the mass predictions and the observed data for all families, if
considering the spin-dependent interaction is ignored in present work. In
spite of deviations about $8\%$ for some states the mass formula
(\ref{MMq}) is confirmed qualitatively at the level of the average deviation
less than $5\%\,$. The best agreement occurs for the $\omega/f$ trajectory for
which the average mass deviation is about $3\allowbreak0$MeV.

In Table VII, we list the slopes given by the formula (\ref{ina}), other
calculations and the analysis of the experimental data cited. \medskip\

\renewcommand\tabcolsep{0.3cm}
\renewcommand{\arraystretch}{1.6}
\renewcommand{\raggedright}{\leftskip=0pt \rightskip=0pt plus 0cm}
\begin{table}[!htbp]
\flushleft
{{\small TABLE VI: The masses (MeV) computed by the formula (\ref{MMq}),
compared to the observed data from the experiment (PDG) \cite{Patrignani:C16}.
The mean squared (MS) errors comparing with the observed masses are shown for
each trajectory.}}%

\begin{tabular}
[c]{cccc}\hline\hline
{\small Traj. }$\chi_{MS}^{2}$(GeV$^{2}$) & $\text{{Mesons }}J^{PC}$ &
$\text{{\small Exp.}}$ & {\small This work}\\\hline
$%
\begin{array}
[c]{r}%
\pi/b\ \ \\
(I=1)
\end{array}%
\begin{array}
[c]{c}%
MS\\
0.00175
\end{array}
$ & $%
\begin{array}
[c]{rr}%
b_{1} & 1^{+-}\\
\pi_{2} & 2^{-+}\\
b_{3} & 3^{+-}\\
\pi_{4} & 4^{-+}\\
b_{5} & 5^{+-}%
\end{array}
$ & $%
\begin{array}
[c]{r}%
1229\\
1672\\
2030\\
2250\\
--
\end{array}
$ & $%
\begin{array}
[c]{r}%
1228\\
1672\\
2003\\
2280\\
2524
\end{array}
$\\\hline
$%
\begin{array}
[c]{r}%
\rho/a\ \\
(I=1)
\end{array}%
\begin{array}
[c]{c}%
MS\\
0.00345
\end{array}
$ & $%
\begin{array}
[c]{rr}%
\rho\  & 1^{--}\\
a_{2} & 2^{++}\\
\rho_{3} & 3^{--}\\
a_{4} & 4^{++}\\
\rho_{5} & 5^{--}\\
a_{6} & 6^{++}\\
\rho_{7} & 7^{--}%
\end{array}
$ & $%
\begin{array}
[c]{r}%
775\\
1318\\
1689\\
1995\\
2330\\
2450\\
--
\end{array}
$ & $%
\begin{array}
[c]{r}%
716\\
1324\\
1706\\
2009\\
2270\\
2503\\
2715
\end{array}
$\\\hline
$%
\begin{array}
[c]{r}%
\eta/h\ \ \\
(I=0)
\end{array}%
\begin{array}
[c]{c}%
MS\\
0.01807
\end{array}
$ & $%
\begin{array}
[c]{rr}%
\eta & 0^{-+}\\
h_{1} & 1^{+-}\\
\eta_{2} & 2^{-+}\\
h_{3} & 3^{+-}\\
\eta_{4} & 4^{-+}\\
h_{5} & 5^{+-}%
\end{array}
$ & $%
\begin{array}
[c]{r}%
548\\
1170\\
1617\\
2025\\
2328\\
-
\end{array}
$ & $%
\begin{array}
[c]{r}%
459\\
1233\\
1673\\
2014\\
2305\\
2563
\end{array}
$\\\hline
$%
\begin{array}
[c]{r}%
\omega/f\ \ \\
(I=0)
\end{array}%
\begin{array}
[c]{c}%
MS\\
0.00806
\end{array}
$ & $%
\begin{array}
[c]{rr}%
\omega & 1^{--}\\
f_{2} & 2^{++}\\
\omega_{3} & 3^{--}\\
f_{4} & 4^{++}\\
\omega_{5} & 5^{--}\\
f_{6} & 6^{++}\\
\omega_{7} & 7^{--}%
\end{array}
$ & $%
\begin{array}
[c]{r}%
783\\
1275\\
1667\\
2018\\
2250\\
2469\\
-
\end{array}
$ & $%
\begin{array}
[c]{r}%
772\\
1315\\
1684\\
1981\\
2239\\
2471\\
2682
\end{array}
$\\\hline
$%
\begin{array}
[c]{r}%
K^{\ast}\ \ \\
(S=-1)
\end{array}%
\begin{array}
[c]{c}%
MS\\
0.01672
\end{array}
$ & $%
\begin{array}
[c]{rr}%
K^{\ast} & 1^{-}\\
K_{2}^{\ast} & 2^{+}\\
K_{3}^{\ast} & 3^{-}\\
K_{4}^{\ast} & 4^{+}\\
K_{5}^{\ast} & 5^{-}\\
K_{6}^{\ast} & 6^{+}%
\end{array}
$ & $%
\begin{array}
[c]{r}%
892\\
1426\\
1776\\
2045\\
2382\\
--
\end{array}
$ & $%
\begin{array}
[c]{r}%
863\\
1430\\
1795\\
2087\\
2339\\
2565
\end{array}
$\\\hline
$%
\begin{array}
[c]{r}%
\phi/f^{\prime}\ \\
(S=0)
\end{array}%
\begin{array}
[c]{c}%
MS\\
0.02890
\end{array}
$ & $%
\begin{array}
[c]{rr}%
\phi\  & 1^{--}\\
f_{2}^{\prime} & 2^{++}\\
\phi_{3} & 3^{--}\\
f_{4}^{\prime} & 4^{++}\\
\phi_{5} & 5^{--}%
\end{array}
$ & $%
\begin{array}
[c]{r}%
1019\\
1525\\
1854\\
2255\\
--
\end{array}
$ & $%
\begin{array}
[c]{r}%
986\\
1538\\
1911\\
2213\\
2475
\end{array}
$\\\hline\hline
\end{tabular}
\vspace{3mm}\medskip\
\end{table}

To answer why the determined value of the confining parameter $a$ in
(\ref{aglo}) is relatively smaller than that in the other quark models cited
in Table V, we would like emphasize that our method to solve the model differs
from that in the most quark models in that it requires the light quark
mass to be quite small (close to the bare mass). As stated in the introduction,
our mass formula is obtained not only by applying the AF method to solve the
model (\ref{SH}) in the relatively large-$N$ case, but also using the
empirical Regge linearity that is confirmed experimentally in the large-$N$
states. The later purpose is fulfilled by quasi-linearizing the quark model
formula for the mass squared around the high excited states. When mapping the
observed masses, the parameters in our model, guided also by the linear fit in
Table III, are mainly fixed so that the behavior of the high excited states is
highlighted, where $M_{q\bar{q}}^{2}\simeq8aL$.

In the most of relativistic quark models cited, however, the parameter setup
crucially relies on the low-lying spectrum in which the quark masses are
heavy, roughly around $200\sim300$MeV. This setup of the quark mass will
violate the linearity of Regge trajectories of the low-lying states, provided
that no further relativistic treatment similar to that in
{\small \cite{GodfreyIsgur:85}} is made to the potential $V$ in (\ref{V})
correspondingly. This can be shown using the semiclassical approximation in
the Section 5. \medskip

\renewcommand\tabcolsep{0.15cm}
\renewcommand{\arraystretch}{1.8}
\renewcommand{\raggedright}{\leftskip=0pt \rightskip=0pt plus 0cm}
\begin{table}[!htbp]
\flushleft
{{\small TABLE VII: The slopes computed by Eq. (\ref{ina}) where }$L$
{\small varies from }$0\ ${\small to }$5\ ${\small and by the linear fit in
Table III. Some other typical predictions cited are also listed for
comparison.}}%

\begin{tabular}
[c]{ccc}\hline\hline
Reference & $\alpha^{\prime}$(GeV$^{-2}$) & Methods\\\hline
This work & $0.47$-$0.75$ & Rel. QM + AFM + Quasi-linearizing\\
\cite{EbertFG:D09} & $0.887/0.839$ & Rel. QM + Pseudo-Potential\\
\cite{SonnenscheinW:JHEP14} & $0.884$ & Massive quark + Rel. string\\
\cite{JohnsonN:PRD79} & $0.88$ & Massive quark + Rel. string\\
This work & $0.832\pm0.059$ & Linear fit (data in PDG 2016)\\
\cite{AnisovichAS:PRD00} & $0.80\pm0.10$ & Linear fit (data in PDG 1998)\\\hline\hline
\end{tabular}
\end{table}

%

%TCIMACRO{\FRAME{ftbpFU}{4.4978in}{3.5854in}{0pt}{\Qcb{\QTR{small}{The mass
%squared }$M^{2}$\QTR{small}{ vs. orbital angular momentum }$L$\QTR{small}{ for
%the nonstrange mesons. Four trajectories are the }$\pi/b$\QTR{small}{
%family(a), the }$\rho/a$\QTR{small}{ family(b), the }$\eta/h$\QTR{small}{
%family(c) and the }$\omega/f$\QTR{small}{ family(d).}}}{}{reggepibf.eps}%
%width 4.4978in;  height 3.5854in;  depth 0pt;  original-width 5.775in;
%original-height 4.5954in;  cropleft "0";  croptop "1";  cropright "1";
%cropbottom "0";  filename 'ReggePibF.eps';file-properties "XNPEU";}} }%
%BeginExpansion
\begin{figure}
[ptb]
\begin{center}
\includegraphics[
width=3in
]%
{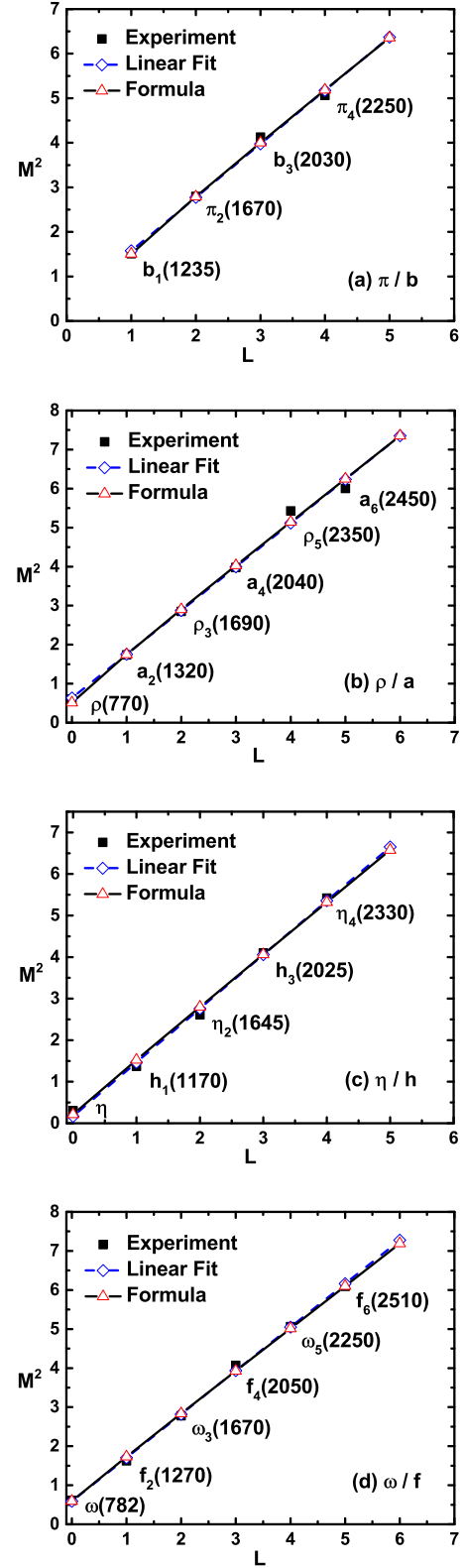}%
\caption{{\protect\small The mass squared }$M^{2}$ {\protect\small  vs. orbital
angular momentum }$L$ {\protect\small  for the nonstrange mesons. Four
trajectories are the }$\pi/b$ {\protect\small  family (a), the }$\rho
/a$ {\protect\small  family (b), the }$\eta/h$ {\protect\small  family (c) and the
}$\omega/f$ {\protect\small  family (d).}}%
\end{center}
\end{figure}
%EndExpansion
%TCIMACRO{\FRAME{ftbpFU}{4.561in}{1.7884in}{0pt}{\Qcb{\QTR{small}{The mass
%squared }$M^{2}$\QTR{small}{ vs. orbital angular momentum }$L$\QTR{small}{ for
%the strange mesons. Two trajectories are the }$K^{\ast}$\QTR{small}{ family
%(a) and the }$\phi/f^{\prime}$\QTR{small}{ family(b)}}}{}{regge-kphi.eps}%
%width 4.561in;  height 1.7884in;  depth 0pt;  original-width 5.775in;
%original-height 2.0409in;  cropleft "0";  croptop "1.0991";  cropright "1";
%cropbottom "0";  filename 'Regge-KPhi.eps';file-properties "XNPEU";}} }%
%BeginExpansion
\begin{figure}
[ptb]
\begin{center}
\includegraphics[
width=3.4in
]%
{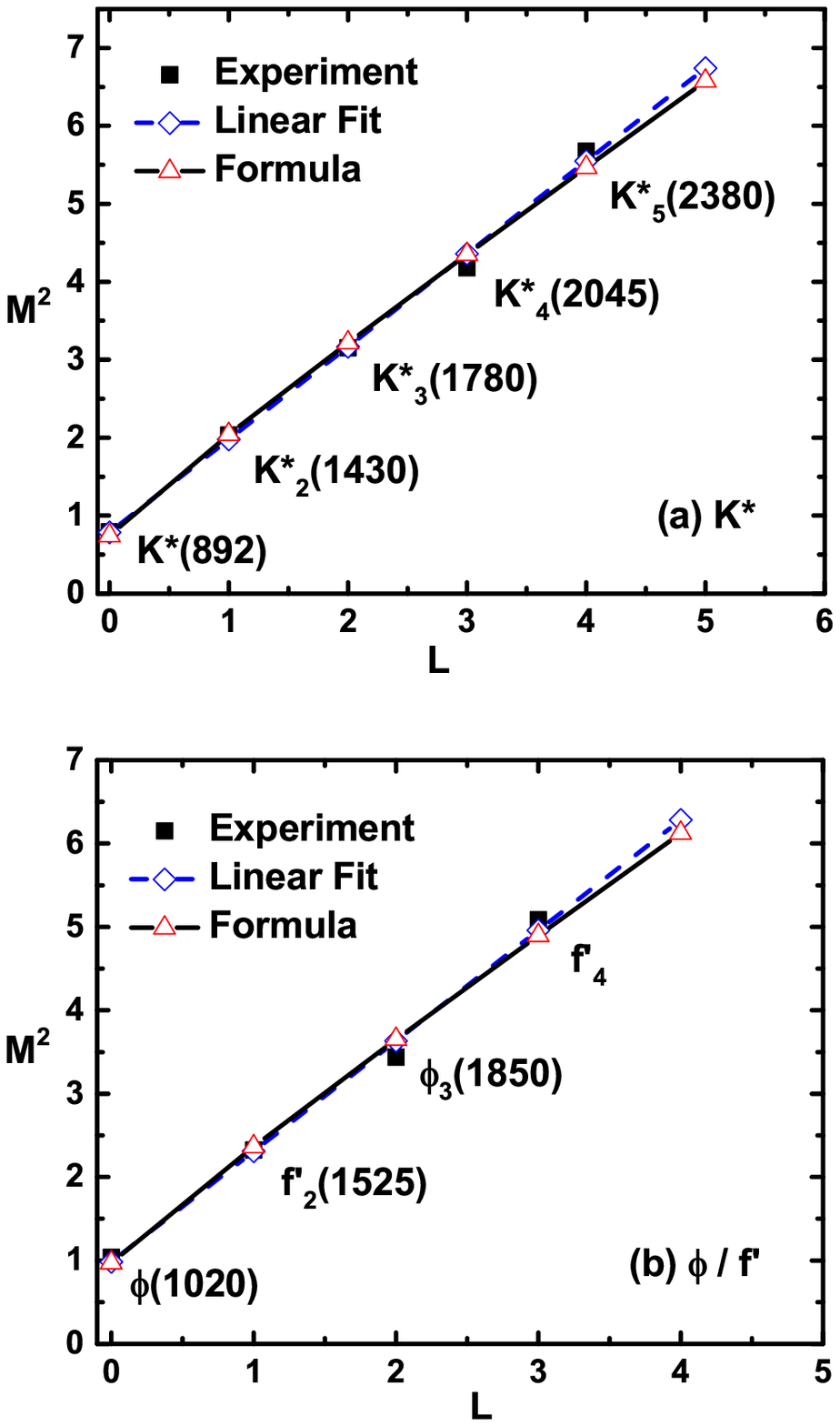}%
\caption{{\protect\small The mass squared }$M^{2}$ {\protect\small  vs. orbital
angular momentum }$L$ {\protect\small  for the strange mesons. Two trajectories
are the }$K^{\ast}$ {\protect\small  family (a) and the }$\phi/f^{\prime}%
$ {\protect\small  family (b).}}%
\end{center}
\end{figure}
%EndExpansion
\medskip

One sees that our Regge-like mass relation (\ref{MMq})
for the orbitally-excited light mesons agrees well with the observed mass
spectrum. Moreover, the relation has a feature that supports a universality
underlying in formation dynamics of the light meson in the following sense:

(i) The quark mass dependence of the light meson spectroscopy is suppressed
doubly by $\bar{m}^{2}$ and ${1/N}$ in the high excited states.

(ii) The confining parameter $a$ and the cutoff $\Lambda$ is nearly same for
all families of the mesons considered, except for the $\eta/h$ trajectory.

We remark that Table IV does not indicate flavor-independence of $V_{0}$
though their dependence on the flavor contents is weak. This is so because
$V_{0}$ in practice accounts for all residual contributions including the
averaged spin-dependent interaction which depends spin nature of mesons
eventually in net value.

Accepting the above, one infers that the members of the $\eta/h$ trajectory
should contain exotic components which makes them exceptional, as seen in
Table IV. The usual mixing $(\sqrt{2}nn-2\bar{s}s)/\sqrt{6}$ for $\eta/h$
trajectory is not adequate for accounting for their abnormal feature.

\section{Semiclassical approximation}\label{sec5}

Before arguing why the quark mass $m_{i}$ should be small in our model, let us
check numerically what value the confining parameter will be equivalent to
when the relativistic correction due to the rotating of QCD string tied to
quarks is taken account into. In the quark model, this correction has been
ignored intrinsically by the potential assumption of the interquark
interaction. Nevertheless, one can check how the data mapping of model makes
up the deficit by comparing the slopes between them. Our model predicts the slope $1/(8a)$ in the
high excited limit of (\ref{ina}), while the rotating string
model predicts the Regge slope $1/(2\pi\sigma)$, where $\sigma$ is the string
tension. The result $a=0.147$ in (\ref{aglo}) is equivalent, under the
following correspondence,
\begin{equation}
\frac{1}{8a}\longleftrightarrow\frac{1}{2\pi\sigma}, \label{map}%
\end{equation}
to the string tension $\sigma=0.187$GeV$^{2}$, which agrees well with that in
the most of relativistic quark models.

While it has been already known in the
past \cite{Dias:Z85-86,CeaNP:D82,CeaNP:D83,Goebel:D90,VeseliO:B96} we would
like to reemphasize the connection between the linear confinement, linear
trajectories and relativistic dynamics \cite{LuchaSG:RP91}, which is helpful
to understand the results in Table II (a,b), Table V and VI. Firstly, we note
that the leading Regge slope follows from the correspondence (classical)
limit. Let us consider the radially-lowest state of (\ref{SH}) but with a
given large $L$ which corresponds to the circular orbital motion of quark at
large $r$ and large $p$. The minimal energy condition $(\partial H/\partial
r)|_{L}\equiv0$ for $H$ in (\ref{SH}) implies that $L|\mathbf{p}|[\mu_{1}%
^{-1}+\mu_{2}^{-1}]=ar^{2}$, with $\mu_{i}=\sqrt{|\mathbf{p}|^{2}+m_{i}^{2}}$.
If one takes $m_{1,2}\rightarrow0$ (the light-light limit), then $\mu
_{1,2}\rightarrow|\mathbf{p}|$ and $|\mathbf{p}|^{2}\rightarrow L^{2}/r^{2}%
$($p_{r}\ll1$ ).$\quad$It follows that
\begin{equation}
H\rightarrow2|\mathbf{p}|+ar\text{, }2L=ar^{2}, \label{HL}%
\end{equation}
which yields (when using $|\mathbf{p}|=L/r$ and eliminating $r$)
\begin{equation}
\alpha_{LL}^{\prime}\text{(}L\gg1\text{)}=\frac{L}{H^{2}}=\frac{1}{8a},
\label{aLL}%
\end{equation}
This is in consistent with (\ref{ina}).

It is of heuristic to "derive" the Regge relation (\ref{aLL}) with the help of
Bohr-like argument for Hydrogen atom. From (\ref{HL}) one has, for the large-$L$ state
$|0L\rangle$,
\begin{equation}
\langle H\rangle_{L}=2a\langle r\rangle_{L},2L=a\langle r^{2}\rangle_{L},
\label{H0L}%
\end{equation}
with $n=0$ assumed. Usage of (\ref{vva}) yields $\langle r\rangle
_{L}=\nu_{N}/a=\sqrt{2L/a}$. It follows that
\begin{equation}
\langle H\rangle_{L}=\sqrt{8aL},\langle r^{2}\rangle_{L}\sim\frac{L}{a},
\label{H8}%
\end{equation}
as required to have (\ref{aLL}).

On the other hand, if we assume $m_{1,2}/M$ is not small, with $M$ the mass of
the quark-antiquark system, a similar analysis that leads to (\ref{WKB})
yields (in the case that $m_{1,2}=m$, $L=0$) a WKB quantization condition
\[
(n+b^{\prime})\pi=\int_{-(M_{c}-m)/a}^{(M_{c}-m)/a}dx\sqrt
{(M_{c}-a|x|)^{2}-m^{2}}.
\]
with $M_{c}=M/2$. It follows that
\begin{equation}
2a\pi(n+2b^{\prime})=M^{2}F_{d}\left(  \frac{2m}{M}\right)  , \label{Mnn}%
\end{equation}%
\begin{equation}
F_{d}(\tau)=\sqrt{1-\tau^{2}}-\tau^{2}\ln\left(  \frac{1+\sqrt{1-\tau^{2}}%
}{\tau}\right)  \label{Fd}%
\end{equation}
where the procedure $n\rightarrow n/2$ was used to remove the reflection
($x\rightarrow-x$) degeneration (double counting of the radial space of
quark motion). One sees from FIG. 5 that the deviation from the linear Regge
trajectory, given by $1-F_{d}(2m/M)$, can be up to $\allowbreak25\%$ for
$m=220$MeV and $\allowbreak40\%$ for $m=300$MeV if choosing $M$ to be mass of
$a_{2}(1318)$ $2^{++}$. This explains why the quark mass $m_{i}$ should be
small in our model, compared to the most of the quark models cited in Table V.
%
%TCIMACRO{\FRAME{ftbpFU}{2.9404in}{2.2525in}{0pt}{\Qcb{$F_{d}$\QTR{small}{ \ as
%a function of }$2m/M$}}{}{fig5-fd.eps}{\special{ language "Scientific Word";
%type "GRAPHIC";  maintain-aspect-ratio TRUE;  display "USEDEF";
%valid_file "F";  width 2.9404in;  height 2.2525in;  depth 0pt;
%original-width 8.4427in;  original-height 6.4565in;  cropleft "0";
%filename 'fig5-Fd.eps';file-properties "XNPEU";}} }%
%BeginExpansion
\begin{figure}
[ptb]
\begin{center}
\includegraphics[
width=3.4in
]%
{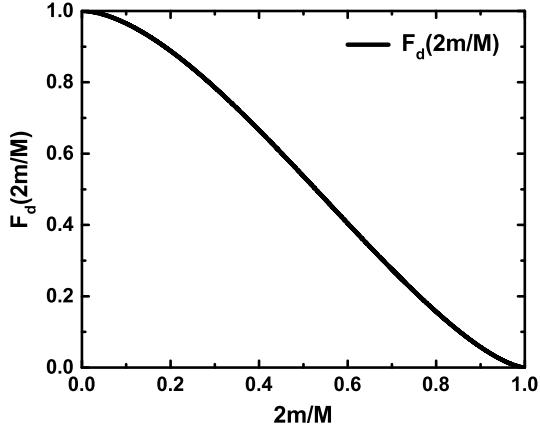}%
\caption{$F_{d}${\protect\small  \ as a function of }$2m/M$.}%
\end{center}
\end{figure}
%EndExpansion

It is of interest to note that the method in Section 2 and 3 can be extended
to the case of heavy-light mesons which consist of a heavy quark and a light
antiquark, though this is not the main issue in this work. Starting again from
the Hamiltonian (\ref{SH}), with $m_{1}=m_{Q}$ being the heavy quark mass and
$m_{2}=m_{q}$ the mass of light antiquark, and assuming $m_{Q}$ to be heavy,
one has, for the Hamiltonian of heavy-light mesons,
\begin{align}
H^{hl}  &  =m_{Q}+\frac{\mathbf{p}^{2}}{2m_{Q}}+\sqrt{\mathbf{p}^{2}+m_{q}%
^{2}}+ar\nonumber\\
&  =m_{Q}+\frac{\mathbf{p}^{2}}{2m_{Q}}+\min_{\mu_{2}}\left\{  \frac
{\mathbf{p}^{2}+m_{q}^{2}}{2\mu_{2}}+\frac{\mu_{2}}{2}\right\} \nonumber\\
&  +\min_{\nu}\left\{  \frac{a^{2}r^{2}}{2\nu}+\frac{\nu}{2}\right\}  ,
\label{H-L}%
\end{align}
where the color-Coulomb term and $V_{0}$ are ignored for simplicity.
Transforming to the center-of-mass system, one has%
\begin{equation}
H^{hl}-m_{Q}=\frac{\mathbf{p}^{2}}{2\mu}+\frac{\mu}{2}\left(  \frac{a^{2}}%
{\mu\nu}\right)  r^{2}+\frac{\mu_{2}+\nu}{2}+\frac{m_{q}^{2}}{2\mu_{2}},
\label{hmQ}%
\end{equation}
with $\mu=\mu_{2}/(1+\mu_{2}/m_{Q})$ the reduced mass of the quarks. The
similar analysis as in section 2 yields%
\begin{equation}
E_{N}(\mu_{2},\nu)=\frac{a_{N}}{\sqrt{\mu\nu}}+\frac{\mu_{2}+\nu}{2}%
+\frac{m_{q}^{2}}{2\mu_{2}}. \label{En2}%
\end{equation}

The minimization of (\ref{En2}) with respect to $\mu_{2}$ and $\nu$ gives
$\mu_{2}=b_{N}^{1/3}[1-\frac{1}{3}b_{N}^{1/3}/m_{Q}]$ and $a_{N}^{2}%
=b_{N}^{1/3}\left[  1-\frac{4}{3}b_{N}^{1/3}/m_{Q}\right]  $ with $b_{N}\equiv
a_{N}^{2}/\nu$. Assuming $m_{Q}\gg1$, one can show
\begin{align*}
\nu^{2}  &  =a_{N}=(a_{N}\nu)^{2/3}=\mu\nu,\\
\mu_{2}  &  =b_{N}^{1/3}=\nu,
\end{align*}
which enables us to rewrite (\ref{En2}) as
\begin{equation}
E_{N}-m_{Q}=2\sqrt{a\left(  N+\frac{3}{2}\right)  }+\frac{m_{q}^{2}}%
{2\sqrt{a_{N}}}, \label{EnQ}%
\end{equation}
or equivalently, as the linear Regge relation $(E_{N}-m_{Q})^{2}=4a\left(
N+3/2\right)  $ in the heavy-light limit. The later relation has an inverse
slope $4a$, being a half of that of the light-meson trajectories at
light-light limit: $\alpha_{HL}^{^{\prime}}=2\alpha_{LL}^{^{\prime}}$. This
feature has been pointed out in
Refs. \cite{JohnsonN:PRD79,VeseliO:B96,Olsson:D1997,Selem:06}. Comparing with the case
of the light mesons, the above argument using the AF method differs in that
only one auxiliary field ($\mu_{2}$) is introduced for the kinematic of light
quark, with the heavy quark treated nonrelativistically. This is in consistent
with the observation \cite{Olsson:D1997} that linear Regge trajectories result
from light quark kinematics and linear confinement.

\section{Summary and concluding remarks}\label{sec6}

Up to date, mesons remain to be the ideal subjects for the study of strong
interactions in the strongly coupled regime. Even though we have a theory of
the strong interactions (QCD), we still know very few about the physical states
of the theory which are crucial to understand QCD eventually. To a large
extent our knowledge of hadron physics relies on phenomenological models, for
instance, the quark models and others. Though successful, the quark models
manifest themselves in various forms, and their predictions can differ
appreciably \cite{GodfreyN:P1999}, in particular, for the excited states. So it
entails constraining of the model predictions by experiment in the case of the
excited states.

For the excited mesons, which can be generated abundantly, the issues such as
whether non-conventional states (exotics) emerge becomes of great interest in
the light sector of hadrons. However, to have hope of distinguishing between
conventional and exotic mesons, it is crucial for us to understand
conventional meson spectroscopy well \cite{GodfreyN:P1999,AmslerT:RP04}.
Great efforts have been made to describe light meson spectrum
\cite{RujulaGG:D75,KangS:D75,BasdevantB:C85,GodfreyIsgur:85,EbertFG:D09} in an
universal way, in which model parameters are assumed to be more or less
universal. These descriptions succeeded remarkably in describing the most observed
states. Meanwhile, question as to whether the universality persists remains to
be of important when new discovered states are considered.

In this work, we addressed the orbitally-excited Regge spectrum of the light
mesons and their universality using relativized quark model combined with
approximated linearity of Regge trajectory. By solving the model with the
auxiliary field method and quasi-linearizing the solution near the asymptotic
limit of the large orbital angular momentum, an new Regge-like mass relation
is proposed for the orbitally excited mesons which supports the universality
that the quark mass dependence of the light meson spectroscopy is suppressed
significantly and the confining parameters $a$ is almost same for all families
except for $\eta/h$-trajectory. The resulted predictions are found to be in
good agreement with the observed data of light mesons.

An explicit expressions for the Regge slope and intercept are obtained and one
mass of the high exciton is predicted for each family in Table VI. We suggest
that the members of the $\eta/h$ trajectory may contain components of exotic.

We have also discussed our results in comparison with the results from the
string (flux-tube) picture of mesons \cite{Nambu:D74,Goto:71} and from the other
quark models. By the way, we presented a semiclassical argument that the
inverse slopes on the radial and angular-momentum Regge trajectories are equal
in the massless limit of quarks, being in consistent with the suggestions in
the literatures.

%\appendix

\vfill
%\newpage

\section*{ACKNOWLEDGMENTS}

D. J is grateful to Atsushi Hosaka and Xiang Liu for many discussions.
D. J is supported by the National Natural Science Foundation of China under
the no. 11565023 and the Feitian Distinguished Professor Program of
Gansu (2014-2016). W. D is supported by
Undergraduate Innovative Ability Program 2018 (Grants No. CX2018B338).

\section*{APPENDIX A}

In order to solve (\ref{AF2}) using the method of homotopic
analysis (HA) \cite{Liao}, we rewrite two nonlinear equations (\ref{Af2}) with
$\bar{m}^{2}=0$ and (\ref{AF2}) in the form of $L_{\nu}(q)=0$ and $NL_{\nu
}(q)=0$, where
\begin{align}
L_{\nu}(q)  &  \equiv\frac{3}{2}-\frac{6a_{N}^{2}}{\nu^{4}},\tag{A-1}\\
NL_{\nu}(q)  &  =\frac{3}{2}+\frac{3ak_{\infty}q^{2}}{N_{\nu}}-\frac
{6a_{N}^{2}}{\nu^{4}\chi_{N}^{2}}-\frac{4ak_{\infty}\nu q^{2}}{N_{\nu}^{3/2}%
}+\frac{16ak_{\infty}a_{N}^{2}q^{2}}{\nu^{3}\chi_{N}^{3}N_{\nu}^{3/2}%
}\nonumber\\
&  +\frac{q^{2}\bar{m}^{2}}{2a_{N}^{2}}\left(  3\nu^{2}\chi_{N}^{2}%
q^{-2}-8ak_{\infty}\frac{\nu^{3}\chi_{N}}{N_{\nu}^{3/2}}\right)  \tag{A-2}%
\end{align}
are two functionals for defining the two equations (\ref{Af2}) and
(\ref{AF2}). The sole difference is that a new and real artificial parameter
$q$ $(0\leq q\leq1)$ is introduced in the above two equations to indicate the
order of smallness (when $N$ becomes large) by following scaling rules (based
on $\nu^{2}\propto N\propto q^{-2}$)%
\begin{align}
\nu &  \rightarrow\nu/q\propto\sqrt{N},a_{N}\rightarrow a_{N}/q^{2}\nonumber\\
N_{\nu}  &  \rightarrow N_{q}/q^{2},N_{q}\equiv(\nu+qa_{L})^{2},\nonumber\\
\chi_{N}  &  \rightarrow\chi_{q}\equiv1+2ak_{\infty}/N_{q}. \tag{A-3}%
\end{align}

The idea of the homotopic analysis (HA) \cite{Liao} is to solve the functional
equation $G(L_{\nu}(q),NL_{\nu}(q),q)$ $=0$ with
\begin{equation}
G(L_{\nu}(q),NL_{\nu}(q),q)\equiv(1-q)L_{\nu}(q)-hqNL_{\nu}(q), \tag{A-4a}%
\end{equation}
or equivalently,
\begin{equation}
(1-q)L_{\nu}(q)=hqNL_{\nu}(q), \tag{A-4b}%
\end{equation}
before solving the nonlinear equation $NL_{\nu}(q=1)=0$ which is (\ref{AF2}).
Here, $h$ is the accelerating factor \cite{Liao} remained to be fixed either by
the platform in the plot window for the characteristic quantity in the model
or comparing with the numerical solution. When $q=0$, $G(L_{\nu}(q),NL_{\nu
}(q),q)=0$ becomes (A-1) while it becomes (\ref{AF2}) when $q=1$. If one
solves (A-4b) (helpful if using the computer) up to the third order of $q$,
one finds
\begin{align}
\nu^{2}  &  =2a_{N}-2ak_{\infty}q^{2}+ak_{\infty}\left(  h+\frac{4a_{L}}%
{\sqrt{2a_{N}}}\right)  q^{3}
\nonumber\\
&  -2\bar{m}^{2}\left(  1+\frac{ak_{\infty}}{2a_{N}%
}(3+4h)q^{2}\right), \tag{A-5}%
\end{align}
which gives (\ref{vsolu}) when putting $q=1$.

\end{document}